\documentstyle[floats,prl,aps]{revtex} 

\input epsf

\def\simlt{\alt}
\def\simgt{\agt}

\begin{document}


\newcommand{\se}{{(0)}}
\newcommand{\ve}{{(1)}}
\newcommand{\te}{{(2)}}
\newcommand{\vepm}{{(\pm 1)}}
\newcommand{\tepm}{{(\pm 2)}}
\newcommand{\po}{{(P)}}
\newcommand{\Spy}[3]{\, {}_{#1}^{\vphantom{#3}} Y_{#2}^{#3}}
\newcommand{\Spyt}[3]{\,{}_{#1}^{\vphantom{1}} \tilde Y_{#2}^{#3*}}
\newcommand{\kap}[3]{\, {}_{#1}^{\vphantom{#3}} \kappa_{#2}^{#3}}
\newcommand{\vertsp}{\vphantom{\displaystyle{\dot a \over a}}}
\newcommand{\grad}{\nabla}

\title{CMB ANISOTROPIES: TOTAL ANGULAR MOMENTUM METHOD}
\author{{Wayne Hu}${}^1$  \&
{Martin White}${}^2$\\
}
\address{${}^1$Institute 
for Advanced Study, School of Natural Sciences\\
Princeton, NJ 08540  \\
${}^2$Enrico Fermi Institute, University of Chicago\\
Chicago, IL 60637
}

\maketitle

\begin{abstract}
A total angular momentum representation simplifies the radiation
transport problem for temperature and polarization anisotropy in
the CMB. Scattering terms couple only the quadrupole moments of the
distributions and each moment corresponds directly to the
observable angular pattern on the sky.  We develop and employ 
these techniques to study the general properties of anisotropy 
generation from scalar, vector and tensor perturbations to the 
metric and the matter, both in the cosmological fluids and
from any seed perturbations (e.g.~defects) that may be present.  
The simpler, more transparent form and derivation of the Boltzmann 
equations brings out the geometric and model-independent aspects of 
temperature and polarization anisotropy formation.  Large angle 
scalar polarization provides a robust means to distinguish between 
isocurvature and adiabatic models for structure formation in principle.
Vector modes have the unique property that the CMB polarization 
is dominated by magnetic type parity at small angles (a factor
of $6$ in power compared with $0$ for the scalars and $8/13$ 
for the tensors) and hence potentially distinguishable independent of 
the model for the seed.  The tensor modes produce a different sign 
from the scalars and vectors for the temperature-polarization 
correlations at large angles. We explore conditions under which one 
perturbation type may dominate over the others including a 
detailed treatment of the photon-baryon fluid before recombination.
\\
\end{abstract}
\hskip 0.5truecm
\tableofcontents
\section{Introduction}

The cosmic microwave background (CMB) is fast becoming the premier 
laboratory for early universe and classical cosmology.  With
the flood of high quality data expected in the coming
years, most notably from the new MAP \cite{MAP} and Planck Surveyor
\cite {Pla}
satellite missions,
it is imperative that theoretical tools for their interpretation
be developed.  The corresponding techniques involved
should be as physically
transparent as
possible so that the implications for cosmology will 
be readily apparent from the data.

Toward this end, we reconsider the general problem of temperature and
polarization anisotropy formation in the CMB.  These anisotropies
arise from gravitational perturbations which separate into
scalar (compressional), vector (vortical), and tensor (gravity
wave) modes.  In previous treatments, the simple underlying geometrical
distinctions and physical processes involved in their appearance
as CMB anisotropies has been obscured by the choice of representation
for the angular distribution of the CMB.
In this paper, we systematically develop a new representation,
the {\it total angular momentum} representation, which puts vector
and tensor modes for the temperature and all polarization modes
on an equal footing with the familiar scalar temperature modes.  
For polarization, this completes and substantially simplifies the
ground-breaking work of \cite{SelZal,KamKosSte}.  Although we
consider only flat geometries here for simplicity, the framework
we establish allows for straightforward generalization
to open geometries \cite{OpenTen,Tom,AbbSch,HSWZ} 
unlike previous treatments.

The central idea of this method is to employ only observable 
quantities, i.e.~those which involve the total angular dependence of
the temperature and polarization distributions.  By applying this
principle from beginning to end, we obtain a substantial simplification
of the radiation transport problem underlying anisotropy formation.
Scattering terms couple only the quadrupole moments of the 
temperature and polarization distributions.  Each moment of the
distribution corresponds to angular moments on the sky
which allows a direct relation between the fundamental scattering
and gravitational sources and the observable anisotropy through
their integral solutions.

We study the means by which gravitational perturbations of
the 
scalar, vector, or tensor type, originating in either the cosmological
fluids or seed sources such as defects,
form temperature and polarization
anisotropies in the CMB.  
As is well established \cite{SelZal,KamKosSte}, scalar perturbations
generate only the so-called electric parity mode of
the polarization.
Here we show that conversely the ratio of magnetic to electric
parity power is a factor of $6$ for vectors, compared with 
$8/13$ for tensors,
{\it independent} of their source. 
Furthermore, the large angle limits of polarization
must obey simple geometrical constraints for its amplitude 
that differ between
scalars, vectors and tensors.   
The sense of the temperature-polarization cross correlation
at large
angle is also determined by geometric considerations which
separate the scalars and vectors from the tensors \cite{CriCouTur}.
These constraints are important
since large-angle polarization unlike large-angle temperature
anisotropies allow one to see directly scales above the 
horizon at last scattering.  Combined with causal constraints,
they provide robust signatures of causal 
isocurvature models for structure
formation such as cosmological defects.  

In \S \ref{sec:normal} we develop the formalism of the total angular
momentum representation and lay the groundwork for the 
geometric interpretation of the radiation transport problem and
its integral solutions. We further establish the relationship
between scalars, vectors, and tensors and the orthogonal angular
modes on the sphere.  In \S \ref{sec:pertevol}, we treat the
radiation transport problem from first principles.  The
total angular momentum representation simplifies both the 
derivation and the form of the evolution equations for the
radiation.  We present the differential form of these equations,
their integral solutions, and their geometric interpretation.
In \S \ref{sec:tight} we specialize the treatment to the 
tight-coupling limit for the photon-baryon fluid before recombination
and show how acoustic waves and vorticity are generated from
metric perturbations and
dissipated through the action of viscosity, polarization 
and heat conduction.  
In \S \ref{sec:seed}, we provide specific examples 
inspired by seeded models such as cosmological defects. 
We trace the full process that transfers seed fluctuations in 
the matter through metric perturbations
to observable anisotropies in the temperature 
and polarization distributions. 

\section{Normal Modes} \label{sec:normal}

In this section, we introduce the total angular momentum 
representation for the normal modes of fluctuations in a 
flat universe that are used to describe the CMB
temperature and  polarization as well as the metric and
matter fluctuations.
This representation greatly simplifies the
derivation and form of the evolution equations for
fluctuations in \S \ref{sec:pertevol}.  In particular, the angular
structure of modes corresponds directly to the angular
distribution of the temperature and polarization, whereas
the radial structure determines how distant sources contribute to
this angular distribution.

The new aspect of this approach is the isolation 
of the {\it total} angular dependence of the modes by combining
the intrinsic angular structure with that of the plane-wave 
spatial dependence. This property implies that the normal
modes correspond directly to angular structures on the sky
as opposed to the commonly employed technique that isolates 
portions of
the {\it intrinsic} angular dependence and hence a
linear combination of observable modes \cite{Pol}.  
Elements of this approach
can be found in earlier works (e.g. \cite{Tom,AbbSch,BonEfs} for
the temperature and \cite{SelZal} for the scalar and
tensor polarization).  We provide here a systematic study of
this technique which also provides for a substantial simplification of
the evolution equations and their integral solution
 in \S \ref{sec:evolution}, including 
the terms involving the radiation transport
of the CMB.  We discuss in detail how the monopole, dipole 
and quadrupole sources that enter into the radiation transport
problem project as anisotropies on the sky today.

Readers not interested in the formal details may skip this section
on first reading
and simply note
that the temperature and polarization distribution is
decomposed into the modes $Y_\ell^m \exp(i\vec{k}\cdot\vec{x})$
and $\Spy{\pm 2}{\ell}{m} \exp(i\vec{k}\cdot\vec{x})$ with
$m=0,\pm 1,\pm 2$ for scalar, vector and tensor metric
perturbations respectively. In this representation, the geometric
distinction between scalar, vector and tensor contributions to
the anisotropies is clear as is the reason why they do not mix.
Here the $\Spy{\pm 2}{\ell}{m}$
are the spin-2 spherical harmonics \cite{Spin} and were introduced 
to the study of CMB polarization by \cite{SelZal}.   The 
radial decompositions 
of the modes $Y_{\ell'}^m j_{\ell'}^{(\ell m)}(kr)$
and $\Spy{\pm 2}{\ell'}{m}[\epsilon_{\ell'}^{(m)}(kr) \pm 
i \beta_{\ell'}^{(m)}(kr)]$ (for $\ell=2$) 
isolate the total angular dependence
by combining the intrinsic and plane wave angular momenta.

\subsection{Angular Modes}

In this section, we derive the basic properties of the angular
modes of the 
temperature and polarization 
distributions that will be useful in \S \ref{sec:pertevol}
to describe their evolution.   In particular, 
the Clebsch-Gordan relation for the addition of angular 
momentum plays a central role in exposing the
simplicity of the total angular momentum representation.

A scalar, or spin-$0$ field on the sky such as the temperature 
can be decomposed into spherical harmonics
$Y_\ell^m$.  Likewise a spin-$s$ field on the sky
can be decomposed into the spin-weighted spherical 
harmonics $\Spy{s}{\ell}{m}$ and a tensor 
constructed out of the basis vectors 
$\hat{e}_\theta \pm i\hat{e}_\phi$,
$\hat{e}_r$
\cite{Spin}.  The basis
for a spin-2 field such as the polarization is
$\Spy{\pm 2}{\ell}{m}{\bf M}_{\pm}$ 
\cite{SelZal,KamKosSte}
where
\begin{equation}
{\bf M}_{\pm} \equiv {1 \over 2}(\hat{e}_\theta \mp i \hat{e}_\phi)
\otimes (\hat{e}_\theta \mp i \hat{e}_\phi) \, ,
\label{eqn:matrixbasis}
\end{equation}
since it transforms under rotations as a $2 \times 2$ symmetric 
traceless tensor. This property is more easily seen through
the relation to the 
Pauli matrices, ${\bf M}_\pm =\sigma_3 \mp i 
\sigma_1$, in spherical coordinates $(\theta,\phi)$.
The spin-$s$ harmonics are expressed in terms of rotation 
matrices\footnote{see e.g.~Sakurai \cite{Sak}, but note that 
our conventions differ
from those of Jackson \cite{Jac} for $Y_\ell^m$ by $(-1)^{m}$.
The correspondence to \cite{KamKosSte} is $\Spy{\pm 2}{\ell}{m}
= [(\ell-2)!/(\ell+2)!]^{1/2}[W_{(\ell m)} \pm i X_{(\ell m)}]$.} 
as
\cite{Spin}
\begin{eqnarray}
\Spy{s}{\ell}{m}(\theta,\phi) &=&
\left( { 2\ell +1  \over 4 \pi} \right)^{1/2}
{\cal D}_{-s,m}^\ell(\phi,\theta,0) \,  \nonumber\\
& = & \left[ {2\ell+1 \over 4\pi} { (\ell+m)!(\ell-m)! 
	\over (\ell+s)! (\ell-s)!} \right]^{1/2} 
	(\sin\theta/2)^{2\ell}
\sum_r \left( 
\begin{array}{c} 
\ell - s \\
r 
\end{array} \right)
\left(
\begin{array}{c} 
\ell + s \\
r + s - m 
\end{array} \right)
		\nonumber\\
&&\qquad \times 
(-1)^{\ell-r-s} e^{im\phi} (\cot\theta/2)^{2r+s-m} .
\label{eqn:spinbasis}
\end{eqnarray}
The rotation matrix ${\cal D}_{-s,m}^{\ell}(\phi,\theta,\psi)
= \sqrt{4\pi/(2\ell+1)} \Spy{s}{\ell}{m}(\theta,\phi)e^{-is\psi}$
represents rotations by the Euler angles $(\phi,\theta,\psi)$.
Since the spin-2 harmonics will be useful in the following
sections, we give their explicit form in Table 1 for $\ell=2$; 
the higher $\ell$ harmonics are related to the 
ordinary spherical harmonics as
\begin{equation}
\Spy{\pm 2}{\ell}{m} = \left[ {(\ell-2)! \over (\ell+2)!} 
\right]^{1/2} \left[ \partial_\theta^2 - {\rm cot}\theta \,
\partial_\theta \pm {2 i \over \sin\theta} (\partial_\theta - {\rm cot}
\theta) \partial_\phi - {1 \over \sin^2\theta} \partial^2_\phi \right]
Y_\ell^m.
\end{equation}

\begin{figure}
\begin{center}
\begin{tabular}{| c | c  | c | }
\hline
$m$ & $Y_2^m$ & $\Spy{2}{2}{m}$  \\ \hline
2 & ${1 \over 4} \sqrt{ 15 \over 2\pi}\, \sin^2\theta\, e^{2i\phi}$ 
  &
	${1 \over 8}\sqrt{ 5 \over \pi}\, (1-\cos\theta)^2 \, e^{2i\phi}$
	$\vertsp$\\ \hline
1 & $\sqrt{15 \over 8\pi} \sin\theta\, \cos\theta \, e^{i\phi}$ 
  &
	${1 \over 4}\sqrt{ 5 \over \pi} \sin\theta\, (1-\cos\theta)\, 
	e^{i\phi}$  $\vertsp$\\ \hline
0 &     ${1 \over 2}\sqrt{5 \over 4\pi}\, (3 \cos^2\theta - 1)$
  &
	${3 \over 4}\sqrt{ 5 \over 6\pi} \, \sin^2\theta$
	 $\vertsp$\\ \hline
-1& $-\sqrt{15 \over 8\pi} \sin\theta\, \cos\theta \, e^{-i\phi}$ 
  &
	${1 \over 4}\sqrt{ 5 \over \pi} \, \sin\theta \,(1+\cos\theta) 
	\, e^{-i\phi} $ 
	 $\vertsp$\\ \hline
-2 & ${1 \over 4} \sqrt{ 15 \over 2\pi}\, \sin^2\theta \, e^{-2i\phi}$ 
   & ${1 \over 8}\sqrt{ 5 \over \pi}\, (1+\cos\theta)^2\, e^{-2i\phi}$
	 $\vertsp$ \\ \hline
\end{tabular}
\\
\vskip 0.5cm
TAB. 1: Quadrupole ($\ell=2$) harmonics for spin-$0$ and $2$.  
\end{center}
\end{figure}

\begin{figure}[t]
\begin{center}
\leavevmode
\epsfxsize=4in \epsfbox{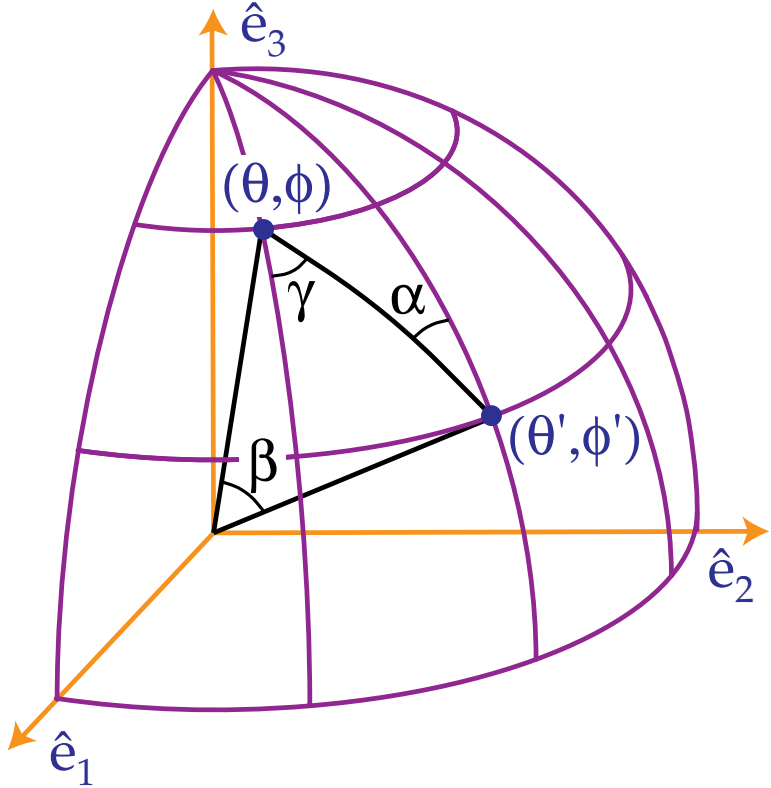} 
\end{center}
\caption{Addition theorem and scattering geometry.  The addition
theorem for spin-$s$ harmonics Eqn.~(\ref{eqn:composition}) is
established by their relation to rotations Eqn.~(\ref{eqn:spinbasis})
and by noting that 
a rotation from $(\theta',\phi')$ through
the origin (pole) to $(\theta,\phi)$ is equivalent to a direct
rotation by the Euler angles $(\alpha,\beta,\gamma)$.  For the
scattering problem of Eqn.~(\ref{eqn:RSR}), these angles
represent the rotation by $\alpha$ from the $\hat{k} = \hat{e}_3$ 
frame to the scattering frame, by the scattering angle $\beta$, and
by $\gamma$ back into the $\hat{k}$ frame.}
\label{fig:scatgeom}
\end{figure}

By virtue of their relation to the rotation 
matrices,
the spin harmonics
satisfy: the compatibility relation
with spherical harmonics,
$\Spy{0}{\ell}{m} = Y_{\ell}^m$;
the conjugation relation $\Spy{s}{\ell}{m*} =  (-1)^{m+s}
\Spy{-s}{\ell}{-m}$; 
the orthonormality relation,
\begin{equation}
\int d\Omega \, \left(\Spy{s}{\ell}{m*}  \right)
             \, \left(\Spy{s}{\ell}{m}   \right)=
        \delta_{\ell,\ell'}
                                      \delta_{m,m'};
\label{eqn:orthogonality}
\end{equation}
the completeness relation,
\begin{equation}
\sum_{\ell,m} \left[ \Spy{s}{\ell}{m*}(\theta,\phi)\right]
              \left[ \Spy{s}{\ell}{m}(\theta',\phi') \right]   =
        \delta(\phi-\phi')\delta(\cos\theta-\cos\theta') \, ;
\label{eqn:completeness}
\end{equation}
the parity relation,
\begin{equation}
\Spy{s}{\ell}{m} \rightarrow (-1)^\ell \Spy{-s}{\ell}{m} ;
\label{eqn:parity}
\end{equation}
the generalized addition relation,
\begin{equation}
\sum_m \left[ \Spy{s_1}{\ell}{m*} (\theta',\phi') \right]
	\left[ \Spy{s_2}{\ell}{m}(\theta,\phi)   \right]
= \sqrt{2\ell+1 \over 4\pi} \left[ \Spy{s_2}{\ell}{-s_1}
	(\beta,\alpha) \right] e^{-i s_2 \gamma} \, ,
\label{eqn:composition}
\end{equation}
which follows from the group multiplication property of
rotation matrices which relates a rotation from $(\theta',\phi')$
through the origin to $(\theta,\phi)$ with a direct rotation
in terms of the Euler angles $(\alpha,\beta,\gamma)$ 
defined in Fig. \ref{fig:scatgeom};
and the Clebsch-Gordan relation,
\begin{eqnarray}
\left(\Spy{s_1}{\ell_1}{m_1} \right)
\left(\Spy{s_2}{\ell_2}{m_2} \right)&=&
{\sqrt{(2\ell_1 + 1)(2\ell_2 + 1) }\over 4\pi} \sum_{\ell,m,s}
\left< \ell_1, \ell_2 ; m_1, m_2 | \ell_1, \ell_2 ; \ell, m \right>
\nonumber\\
&&\times
\left< \ell_1, \ell_2 ; -s_1, -s_2 | \ell_1, \ell_2 ; \ell, -s \right>
\sqrt{4\pi \over 2\ell+1}\, \left( \Spy{s}{\ell}{m}\right)  \, .
\label{eqn:ClebschGordan}
\end{eqnarray}

It is worthwhile to examine the implications of these properties.
Note that the orthogonality and completeness relations 
Eqns.~(\ref{eqn:orthogonality}) and (\ref{eqn:completeness}) do not 
extend to different spin states.  Orthogonality between $s=\pm 2$
states is established by the Pauli basis of 
Eqn.~(\ref{eqn:matrixbasis}) ${\bf M}_{\pm}^* 
{\bf M}_{\pm}^{\vphantom{*}} =
{\bf 1}$ and ${\bf M}_\pm^* {\bf M}_\mp^{\vphantom{*}} = {\bf 0}$. 
The parity equation (\ref{eqn:parity}) tells us that the 
spin flips under a parity transformation so that unlike 
the $s=0$ spherical harmonics, the higher spin harmonics are not 
parity eigenstates.  Orthonormal parity states can be constructed
as \cite{SelZal,KamKosSte}
\begin{equation}
{1 \over 2}[\Spy{2}{\ell}{m}{\bf M}_+ \pm \Spy{-2}{\ell}{m}{\bf M}_-]
\, ,
\label{eqn:parityeigenstates}
\end{equation} 
which have ``electric'' $(-1)^\ell$ and ``magnetic'' 
$(-1)^{\ell+1}$ type parity for the
$(\pm)$ states respectively.
We shall see in \S \ref{sec:evolution} that the polarization 
evolution naturally
separates into parity eigenstates.  The addition property will
be useful in relating the scattering angle to coordinates
on the sphere in \S \ref{sec:transport}.  Finally 
the Clebsch-Gordan relation Eqn.~(\ref{eqn:ClebschGordan}) is 
central to the following discussion and will be used to  derive
the total angular momentum representation in 
\S \ref{sec:radial} and evolution
equations for angular moments of the radiation \S \ref{sec:evolution}.

\subsection{Radial Modes} \label{sec:radial}

We now complete the formalism needed to describe the
temperature and polarization fields by adding a spatial
dependence to the modes. 
By further separating the {\it radial} dependence
of the modes, we gain insight on their full angular structure.
This decomposition will be useful in
constructing the formal integral solutions of the perturbation 
equations in \S \ref{sec:evolution}.  We begin with its derivation
and then proceed to its geometric interpretation.

\subsubsection{Derivation}

The temperature and polarization distribution of the radiation
is in general a function of both spatial position $\vec{x}$ and angle
$\vec{n}$ defining the propagation direction.
In flat space, we know that
plane waves form a complete basis for the spatial dependence.
Thus a spin-$0$ field like the temperature may be expanded in
\begin{equation}
G_\ell^m  =  (-i)^{\ell} \sqrt{ 4\pi \over 2\ell+1} Y_\ell^m(\hat{n})
        \exp(i \vec{k} \cdot \vec{x}) \, ,
\label{eqn:temperaturebasis}
\end{equation}
where the normalization is chosen to agree with the standard
Legendre polynomial conventions for $m=0$.
Likewise a spin-$2$ field like the polarization may be expanded in
\begin{equation}
{}_{\pm 2}G_\ell^m  =
        (-i)^\ell \sqrt{ {4\pi \over 2\ell+1}}
        [\Spy{\pm 2}{\ell}{m}(\hat{n})] \exp(i\vec{k} \cdot \vec{x})\, .  
\label{eqn:polarizationbasis}
\end{equation}

\begin{figure}[t]
\begin{center}
\leavevmode
\epsfxsize=4in \epsfbox{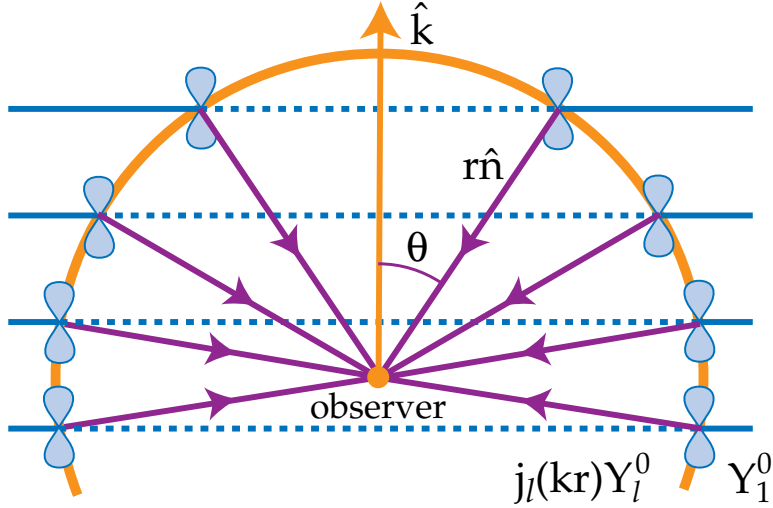} 
\end{center}
\caption{Projection effects.  A plane wave $\exp(i \vec{k} \cdot
\vec{x})$ can be decomposed into $j_\ell(kr) Y_\ell^0$ and
hence carries an ``orbital'' angular dependence.   A 
plane wave source at distance $r$ thus contributes
angular power to $\ell \approx kr$ at $\theta=\pi/2$ but also
to larger angles $\ell \ll kr$ at $\theta=0$ which is 
encapsulated into the structure of $j_\ell$
(see Fig.~\ref{fig:radialtemp}).  If the source has an intrinsic
angular dependence, the distribution of power is altered.  For
an aligned dipole $Y_1^0 \propto \cos\theta$ 
(`figure 8's) power at $\theta=\pi/2$ or
$\ell \approx kr$ is suppressed. These arguments are generalized
for other intrinsic angular dependences in the text. }
\label{fig:diproj}
\end{figure}

The plane wave itself also carries an angular dependence of course,
\begin{equation}
\exp( i \vec{k} \cdot \vec{x} ) 
       =  \sum_\ell (-i)^\ell \sqrt{4\pi(2\ell+1)} j_\ell(kr) 
	Y_\ell^0(\hat{n}) \, ,
\label{eqn:planewave}
\end{equation}
where $\hat{e}_3 = \hat{k}$ and
$\vec{x}
 = - r \hat{n}$ (see Fig.~\ref{fig:diproj}).  
The sign convention for the direction is opposite
to direction on the sky to be in accord with the direction of
propagation of the radiation to the observer.  Thus the extra factor
of $(-1)^\ell$ comes from the parity relation Eqn.~(\ref{eqn:parity}). 

\begin{figure}[t]
\begin{center}
\leavevmode
\epsfxsize=3.5in \epsfbox{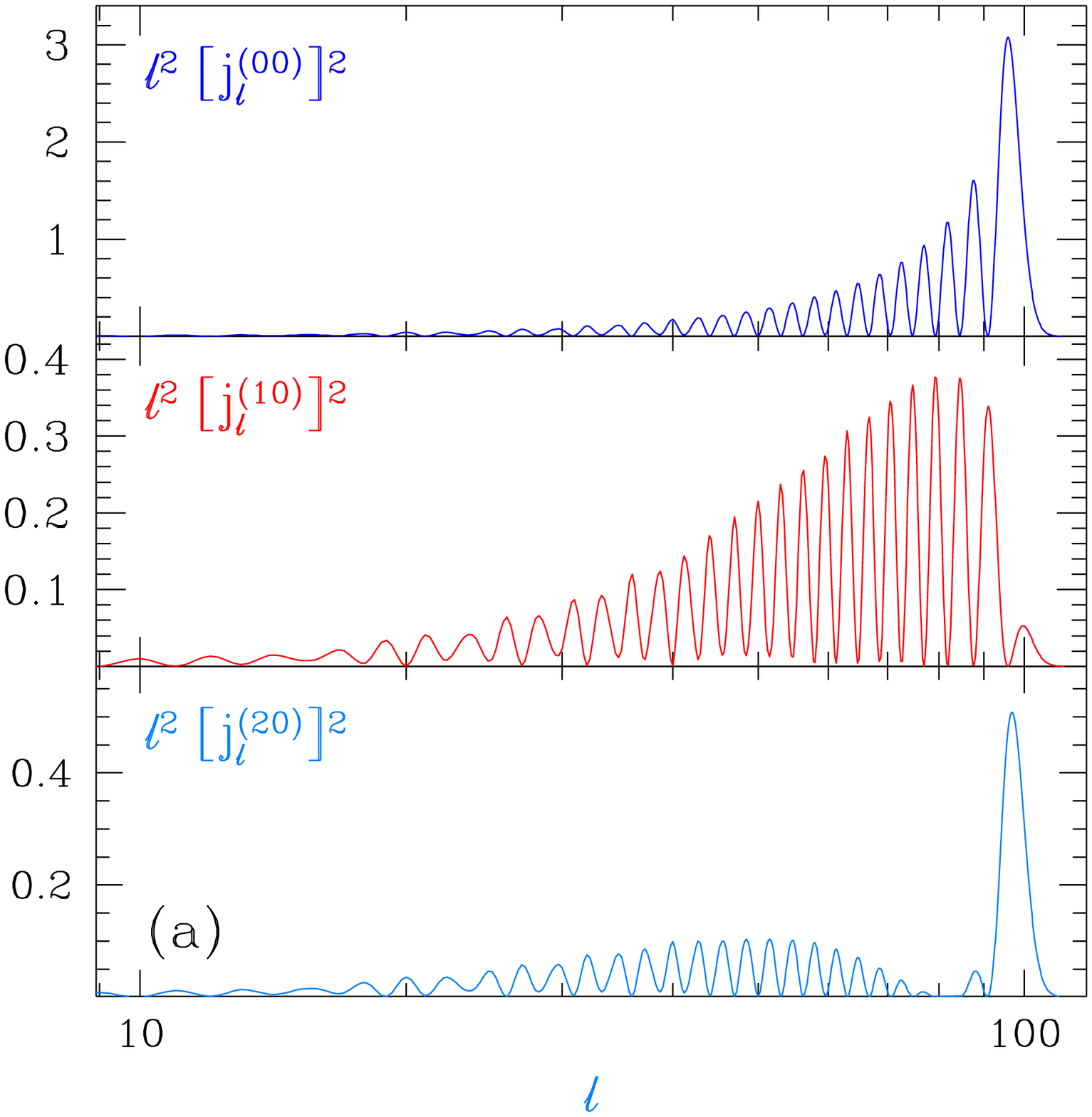} 
\epsfxsize=3.5in \epsfbox{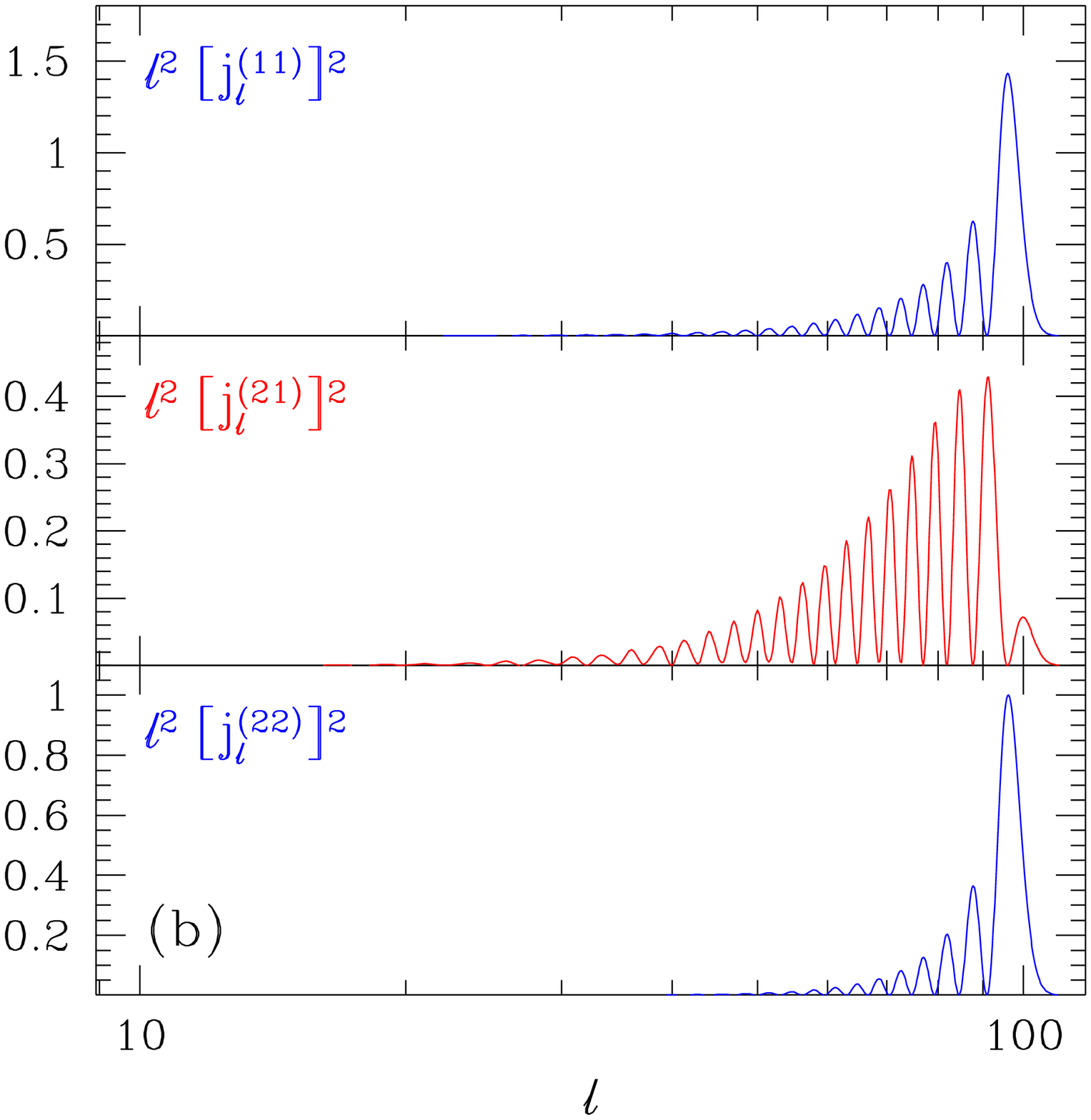} 
\end{center}
\caption{Radial spin-0 (temperature) modes.  
The angular power in a plane wave (left panel, top) 
is modified due to the intrinsic
angular structure of the source as discussed in the text. 
The left panel corresponds to the power in  scalar ($m=0$) 
monopole $G_0^0$, dipole $G_1^0$, and quadrupole $G_2^0$
sources (top to bottom);  
the right panel to that in vector ($m=1$) dipole $G_1^{\pm 1}$
and quadrupole $G_2^{\pm 1}$
sources and a tensor ($m=2$) 
quadrupole $G_2^{\pm 2}$ source (top to bottom).  
Note the differences
in how sharply peaked the power is at $\ell \approx kr$ and
how fast power falls as $\ell \ll kr$.
The argument of the radial functions $kr=100$ here. }
\label{fig:radialtemp}
\end{figure}

The separation of the mode functions into an intrinsic angular
dependence and plane-wave spatial dependence is essentially
a division into spin ($\Spy{s}{\ell'}{m}$) and orbital 
($Y_{\ell}^0$) angular momentum.  Since only
the total angular dependence is observable, it is instructive
to employ the Clebsch-Gordan relation of 
Eqn.~(\ref{eqn:ClebschGordan}) to add the angular momenta.
In general this couples the states between
$|\ell -\ell'|$ and $\ell+\ell'$.  Correspondingly
a state of definite total $\ell$ will correspond to a weighted
sum of $j_{|\ell-\ell'|}$ to $j_{\ell + \ell'}$ in its radial
dependence.  This can be reexpressed in terms of 
the $j_\ell$ using the recursion
relations of spherical Bessel functions,
\begin{eqnarray}
{j_\ell(x) \over x} &=& {1 \over 2\ell+1} 
	[ j_{\ell-1}(x) + j_{\ell+1}(x)] \, , \nonumber\\
j_\ell'(x)  &=& {1 \over 2\ell+1} [ \ell j_{\ell-1}(x) - 
	(\ell +1) j_{\ell+1}(x) ] \,.
\end{eqnarray}
We can then rewrite
\begin{equation}
G_{\ell'}^{m} 
        =   \sum_\ell (-i)^\ell \sqrt{4\pi(2\ell+1)} \,
	  j_\ell^{(\ell' m)}(kr) \, Y_\ell^{m}(\hat {n})  \, ,
\label{eqn:radialtemp}
\end{equation}
where the lowest $(\ell',m)$ radial functions are
\begin{equation}
\begin{array}{lll}
j_\ell^{(00)}(x) = j_\ell(x)  \, ,\qquad &
j_\ell^{(10)}(x) = j_\ell'(x) \, ,\qquad &
j_\ell^{(20)}(x) = {1 \over 2} [3 j_\ell''(x) + j_\ell(x) ] \, ,
	\vphantom{\Bigg[}\\
			    \qquad&
j_\ell^{(11)}(x) = \displaystyle{ 
	\sqrt{\ell(\ell+1) \over 2}\, {j_\ell(x) \over x} }\, ,
			    \qquad&
j_\ell^{(21)}(x) = \displaystyle{ 
	\sqrt{3\ell(\ell+1) \over 2}\, \left({j_\ell(x) \over x}
	\right)' }       \, ,
	\vphantom{\Bigg[} \\
			    \qquad&
			    \qquad&
j_\ell^{(22)}(x) = \displaystyle{ 
	\sqrt{{3 \over 8} {(\ell+2)! \over (\ell-2)!}} \,
		{j_\ell(x) \over x^2} } 
	\vphantom{\Bigg[}, 
\label{eqn:jdef}
\end{array}
\end{equation}
with primes representing derivatives with respect to the argument
of the radial function $x=kr$.
These modes are shown in Fig.~\ref{fig:radialtemp}.

\begin{figure}[t]
\begin{center}
\leavevmode
\epsfxsize=4.0in \epsfbox{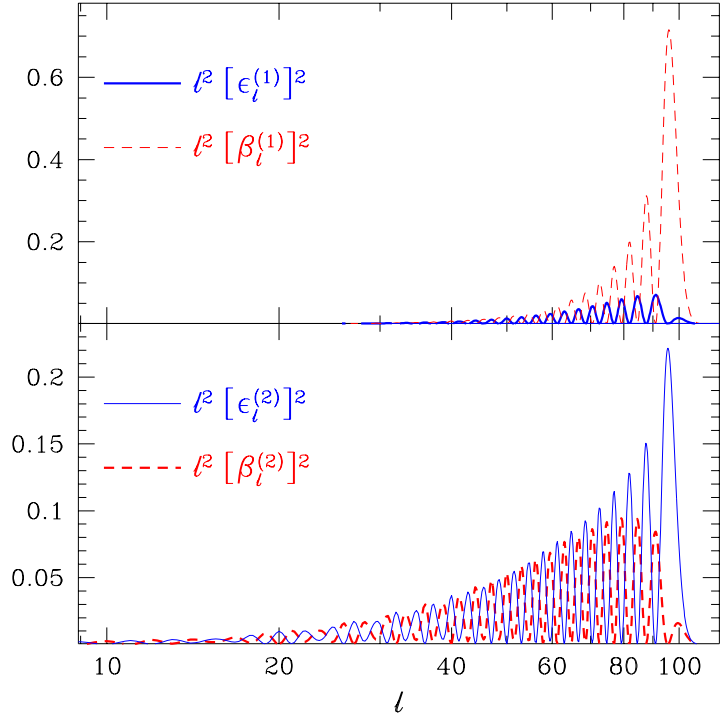} 
\end{center}
\caption{Radial spin-2 (polarization) modes.  Displayed is 
the angular power in a plane-wave spin-2 source.  The top panel
shows that vector ($m=1$, upper panel)
sources are dominated by $B$-parity
contributions, whereas tensor ($m=2$, lower panel) sources have 
comparable but less power in the $B$-parity.  Note that the power is strongly
peaked at $\ell = kr$ for the $B$-parity vectors and $E$-parity
tensors.
The argument of the radial functions $kr=100$ here.}
\label{fig:radialpol}
\end{figure}

Similarly for the spin $\pm 2$ functions with $m>0$ (see 
Fig.~\ref{fig:radialpol}),
\begin{equation}
{}_{\pm 2} G_2^m 
	 =  \sum_\ell (-i)^\ell \sqrt{4\pi(2\ell+1)} 
	[\epsilon_\ell^{(m)}(kr) \pm i\beta_\ell^{(m)}(kr)] \, 
	\Spy{\pm 2}{\ell}{m}(\hat{n}) \, ,
\label{eqn:radialpol}
\end{equation}
where
\begin{eqnarray}
\epsilon^{(0)}_\ell(x) & = &\sqrt{{3 \over 8 }
	{(\ell+2)! \over (\ell-2)!}} {j_\ell(x) \over x^2} \, ,
	\vertsp\nonumber\\
\epsilon^\ve_\ell(x) &=& {1 \over 2} \sqrt{(\ell-1)(\ell+2)}
        \left[ {j_\ell(x) \over  x^2} + {j_\ell'(x) \over  x}
        \right] \, , \nonumber\\
\epsilon^\te_\ell(x) &=& {1 \over 4} \left[ -j_\ell(x)
        + j_\ell''(x) + 2{j_\ell(x) \over x^2} +
        4{j_\ell'(x) \over x} \right]  \, ,
\label{eqn:epsilon}
\end{eqnarray}
which corresponds to the $\ell'=\ell,\ell\pm 2$ coupling and
\begin{eqnarray}
\beta^{(0)}_\ell(x) &=& 0\, , \nonumber\\
\beta^{(1)}_\ell(x) &=& {1 \over 2} \sqrt{(\ell-1)(\ell+2)}
        {j_\ell(x) \over x} \, , \nonumber\\
\beta^{(2)}_\ell(x) &=& {1 \over 2} \left[ j_\ell'(x)
        + 2 {j_\ell(x) \over x} \right] \, ,
\label{eqn:beta}
\end{eqnarray}
which corresponds to the $\ell' = \ell \pm 1$ coupling.
The corresponding relation for negative $m$ involves a reversal
in sign of the $\beta$-functions 
\begin{eqnarray}
\epsilon_\ell^{(-m)} &  =  & \epsilon_\ell^{(m)}  ,\nonumber\\
\beta_\ell^{(-m)} &  =  & -\beta_\ell^{(m)} .
\end{eqnarray}
These functions are plotted in Fig. \ref{fig:radialpol}.
Note that $\epsilon^{(0)}_\ell = j^{(2)}_\ell$ is displayed
in Fig.~\ref{fig:radialtemp}.  

\subsubsection{Interpretation}

The structure of these functions is readily apparent from
geometrical considerations.  A single plane wave contributes
to a range of angular scales from $\ell \approx kr$ at 
$\theta=\pi/2$ to larger angles $\ell \ll kr$
as $\theta \rightarrow (0,\pi)$, where $\hat{k} \cdot \hat{n} 
= \cos\theta$ (see Fig.~\ref{fig:scatgeom}).  
The power in $\ell$ of a
single plane wave shown in Fig.~\ref{fig:radialtemp}(a) (top panel) 
drops to zero $\ell \simgt kr$, has
a concentration of power around $\ell = kr$ and an extended
low amplitude tail to $\ell \simlt kr$.  

Now if the plane
wave is multiplied by an intrinsic angular dependence,
the projected power changes.  
The key to understanding this effect is to note that
the intrinsic angular behavior is related to power in $\ell$
as
\begin{eqnarray}
\theta \rightarrow (0,\pi) \qquad & \Longleftrightarrow & \qquad \ell 
\ll kr \, , 
	\nonumber\\
\theta \rightarrow \pi/2 \qquad & \Longleftrightarrow & \qquad \ell 
\approx kr \, .
\end{eqnarray}
Thus factors of $\sin\theta$ in the intrinsic angular 
dependence suppress power at $\ell \ll kr$ (``aliasing suppression''),
whereas factors of $\cos\theta$ suppress power at $\ell \approx kr$
(``projection suppression'').
Let us consider first a $m=0$ dipole contribution $Y_1^0 \propto
\cos\theta$ (see Fig.~\ref{fig:diproj}).  
The $\cos\theta$ dependence suppresses
power in $j_\ell^{(10)}$ at
the peak in the plane-wave spectrum $\ell \approx kr$ 
(compare Fig.~\ref{fig:radialtemp}(a) top
and middle panels).  The remaining power is broadly distributed
for $\ell \simlt kr$.  
The same reasoning applies for $Y_2^0$ quadrupole sources which
have an intrinsic angular dependence of $3\cos^2\theta - 1$.  
Now the minimum falls at $\theta = \cos^{-1}(1/\sqrt{3})$
causing the double peaked form of the power in $j_\ell^{(20)}$
shown in 
Fig.~\ref{fig:radialtemp}(a) (bottom panel).  
This series can be continued
to higher $G_\ell^0$ and such techniques have been 
used in the free streaming limit for temperature anisotropies
\cite{BonEfs}.

Similarly, the structures of $j_\ell^{(11)}$, $j_\ell^{(21)}$
and $j_\ell^{(22)}$ are apparent from
the intrinsic angular dependences
of the $G_1^1$, $G_2^1$ and $G_2^2$ sources,
\begin{equation}
Y_1^1 \propto \sin\theta e^{i\phi}\, ,  \qquad
Y_2^1 \propto \sin\theta\cos\theta e^{i\phi} \, , \qquad
Y_2^2 \propto \sin^2\theta e^{2i\phi}\, ,
\end{equation} 
respectively.
The $\sin\theta$ factors imply that 
as $m$ increases, low $\ell$ power in the source decreases
(compare Fig.~\ref{fig:radialtemp}(a,b) top panels). $G_2^1$
suffers a further suppression at $\theta = \pi/2$ $(\ell \approx
kr) $ from its $\cos\theta$ factor.

There are two interesting consequences of this behavior.
The sharpness of the radial function around $\ell=kr$ quantifies
how faithfully features in the $k$-space spectrum are preserved
in $\ell$-space.
If all else is equal, this faithfulness increases with $|m|$ for
$G_{|m|}^m$
due to aliasing suppression from $\sin^m\theta.$  
On the other hand, features in $G_{|m|+1}^m$ are
washed out in comparison due projection suppression from the
$\cos\theta$ factor.
 
Secondly, even if there are no contributions from
long wavelength sources with $k \ll \ell/r$, there will still be 
large angle anisotropies
at $\ell \ll kr$ which scale as 
\begin{equation}
[\ell j_\ell^{(\ell'm)}]^2 \propto \ell^{2+2|m|}.
\label{eqn:lowljaliasing}
\end{equation}
This scaling puts an upper bound on how 
steeply the power can rise with $\ell$ that {\it increases} with
$|m|$ and hence a lower bound
on the amount of large relative to small 
angle power that {\it decreases} with $|m|$.

The same arguments apply to the spin-$2$ functions with the 
added complication of the appearance of two radial functions
$\epsilon_\ell$ and $\beta_\ell$.  
The addition of spin-2 angular momenta introduces
a  
$\beta$-contribution from $e^{im\phi}$
except for $m=0$.  For $m = \pm 1$, the $\beta$-contribution strongly 
dominates over the $\epsilon$-contributions;
whereas for $m = \pm 2$, $\epsilon$-contributions are slightly
larger than $\beta$-contributions (see Fig.~\ref{fig:radialpol}).
The ratios reach the asymptotic values of
\begin{equation}
{\sum_\ell [\ell \beta_\ell^{(m)} ]^2  
\over
 \sum_\ell [\ell \epsilon_\ell^{(m)} ]^2 } 
\approx \cases{
	6, & $m= \pm 1,$ \cr
	8/13, & $m= \pm 2,$ \cr}
\label{eqn:ebratio}
\end{equation}
for fixed $kr \gg 1$.  
These considerations 
are closely related to the parity of the multipole expansion.
Although the orbital angular momentum does not mix states of
different spin, it does mix states of different parity 
since the plane wave itself does not have definite parity. 
A state with ``electric'' parity in the intrinsic angular
dependence (see Eqn.~\ref{eqn:parityeigenstates}) becomes
\begin{eqnarray}
{}_{2}^{\vphantom{2}} 
	G_2^m {\bf M}_+ + 
{}_{-2}^{\vphantom{2}} G_2^m {\bf M}_- & = &
	\sum_\ell (-i)^\ell \sqrt{4\pi(2\ell+1)}\Big\{ \epsilon_\ell^{(m)}
	[\Spy{2}{\ell}{m}{\bf M}_+ + \Spy{-2}{\ell}{m}{\bf M}_-] 
	\nonumber\\
&  & \qquad
	+ i\beta_\ell^{(m)}
      [\Spy{2}{\ell}{m}{\bf M}_+ - \Spy{-2}{\ell}{m}{\bf M}_-] \Big\}.
\label{eqn:ebparity}
\end{eqnarray}
Thus the addition of angular momentum of 
the plane wave generates ``magnetic'' $B$-type 
parity of amplitude $\beta_\ell$ 
out of an intrinsically ``electric'' $E$-type source as
well as
$E$-type parity of amplitude
$\epsilon_\ell$.
Thus the behavior of the two radial functions has
significant consequences for the polarization calculation in 
\S \ref{sec:evolution} and implies that B-parity
polarization is absent for scalars, dominant for vectors,
and comparable to but slightly smaller than the E-parity 
for tensors. 

\begin{figure}[t]
\begin{center}
\leavevmode
\epsfxsize=4.0in \epsfbox{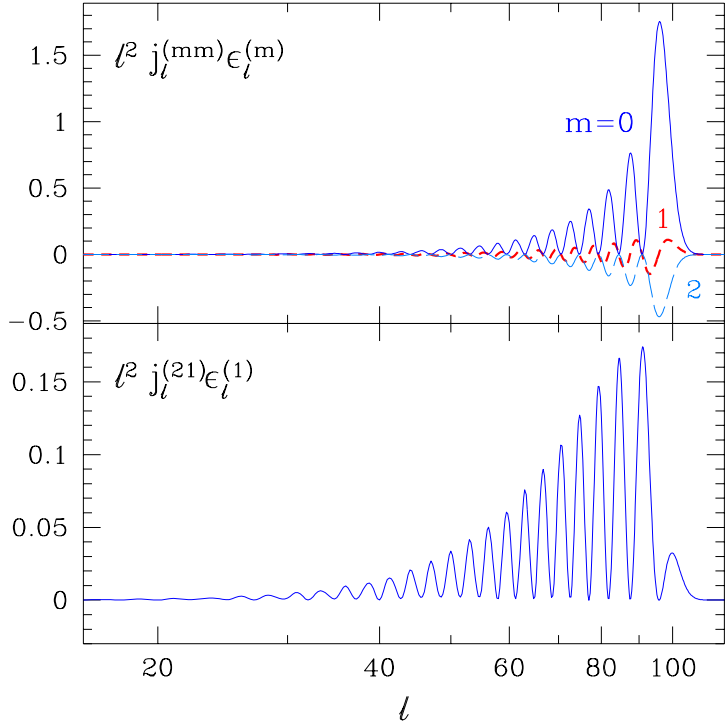} 
\end{center}
\caption{Spin-0 $\times$ Spin-2 (temperature $\times$ polarization)
modes.  Displayed is 
the cross angular power in plane wave spin-0 and spin-2 sources.  
The top panel shows that a scalar monopole ($m=0$) source correlates
with a scalar spin-2 (polarization quadrupole) source whereas
the tensor quadrupole ($m=2$) anticorrelates with a tensor
spin-2 source.  Vector dipole ($m=1$) sources oscillate in
their correlation with vector spin-2 sources and contribute
negligible once modes are superimposed. One must go to
vector quadrupole sources (lower panel) for a strong correlation.
The argument of the radial functions $kr=100$ here.}
\label{fig:radialcross}
\end{figure}

Now let us consider the low $\ell \ll kr$ tail
of the spin-$2$ radial functions.
Unlike the 
spin-0 projection, the spin-2 projection allows increasingly 
more power at $\theta \rightarrow 0$ and/or $\pi$,
i.e.~$\ell \ll kr$, as
$|m|$ increases (see Table 1 and note the factors of $\sin\theta$).  
In this limit, the power
in a plane wave fluctuation goes as 
\begin{equation}
[\ell \epsilon_\ell^{(m)}]^2 \propto \ell^{6 - 2|m|},  \qquad 
[\ell \beta_\ell^{(m)}]^2 \propto \ell^{6 - 2|m|}.
\label{eqn:lowlaliasing}
\end{equation}
Comparing these expressions with Eqn.~(\ref{eqn:lowljaliasing}),
we note that the spin-$0$ and spin-$2$ 
functions have an opposite dependence on $m$.
The consequence is that the relative power in large vs. small
angle polarization 
tends to decrease from the $m=2$ tensors to the $m=0$ scalars.

Finally it is interesting to consider the cross power between
spin-$0$ and spin-$2$ sources because it will be used
to represent the temperature-polarization cross correlation.
Again interesting
geometric effects can be identified
(see Fig.~\ref{fig:radialcross}).  
For $m=0$, the power in $j_\ell^{(00)} \epsilon_\ell^{(0)}$
correlates 
(Fig.~\ref{fig:radialcross}, top panel solid line, positive definite);
for $m= 1$, $j_\ell^{(11)}\epsilon_\ell^{(1)}$ 
oscillates (short dashed line)
and for $m=2$, 
$j_\ell^{(22)} \epsilon_\ell^{(2)}$ anticorrelates  (long dashed line,
negative definite).  
The cross power involves only $\epsilon_\ell^{(m)} j_\ell^{(\ell'm)}$
due to the opposite parity of the $\beta_\ell^{(m)}$ modes.

These properties will become important in \S 
\ref{sec:pertevol} and \ref{sec:viscosity}
and translates into cross power contributions with
{\it opposite} sign between the scalar monopole temperature cross 
polarization sources
and tensor quadrupole temperature cross polarization sources
\cite{CriCouTur}.  
Vector dipole temperature and polarization 
sources do not contribute strongly to the cross power since
correlations and anticorrelations in $j_\ell^{(11)} \epsilon_\ell^{(1)}$
will cancel when modes are
superimposed.  The same is true of the scalar dipole 
temperature cross polarization $j_\ell^{(10)} \epsilon_\ell^\se$
as is apparent from
Figs.~\ref{fig:radialtemp} and \ref{fig:radialpol}. 
The vector cross power is dominated by
quadrupole temperature and polarization sources
$j_\ell^{(21)}\epsilon_\ell^{(1)}$
(Fig.~\ref{fig:radialcross} lower panel).

\subsection{Perturbation Classification} \label{sec:classification}

As is well known (see e.g.~\cite{AbbSch,KodSas}), a 
general symmetric tensor such as the metric and stress-energy
perturbations can be separated into scalar, vector
and tensor pieces through their coordinate transformation 
properties.  We now review the properties of their
normal modes so that they
may be related to those of the radiation.  We find that
the $m=0,\pm 1,\pm 2$ modes of the radiation couple to the
scalar, vector and tensor modes of the metric. Although we 
consider flat geometries here, we preserve
a covariant notation that ensures straightforward generalization
to open geometries through the replacement of $\delta_{ij}$ 
with the curved three metric
and ordinary derivatives with covariant derivatives \cite{Tom,AbbSch}.

\subsubsection{Scalar Perturbations}

Scalar perturbations in a flat universe are represented by
plane waves
$Q^\se = \exp( i \vec{k} \cdot \vec{x} )$,
which are the eigenfunctions
of the Laplacian operator
\begin{equation}
\nabla^2 Q^\se = -k^2 Q^\se  , 
\label{eqn:scalareigen}
\end{equation}
and their spatial derivatives.  For example,
vector and symmetric tensor quantities such as velocities and 
stresses based on scalar perturbations
can be constructed as
\begin{equation}
Q_i^\se = - k^{-1} \nabla_i Q^\se, \qquad Q_{ij}^\se = 
	[k^{-2} \nabla_i \nabla_j + {1 \over 3} \delta_{ij}] Q^\se .
\end{equation}
Since $\vec{\nabla} \times \vec{Q}^\se = 0$, velocity fields
based on scalar perturbations are irrotational.
Notice that $Q^\se = G_0^0$, ${n}^i Q_i^\se = G_1^0$ and 
${n}^i {n}^j Q_{ij}^\se \propto G_{2}^0$,
where the coordinate system is defined by $\hat{e}_3 = \hat{k}$.
From the orthogonality of the spherical harmonics, it follows
that scalars generate only $m=0$ fluctuations in the radiation.

\subsubsection{Vector Perturbations}

Vector perturbations can be decomposed into harmonic modes
$Q^\vepm_i$ of the Laplacian in the same manner as the scalars,
\begin{equation}
\nabla^2 Q^\vepm_{i} = -k^2 Q^\vepm_{i} ,
\label{eqn:vectoreigen}
\end{equation}
which satisfy a divergenceless condition
\begin{equation}
{\nabla}^i {Q}_i^\vepm = 0 \, .
\end{equation}
A 
velocity field based on vector perturbations thus represents vorticity,
whereas scalar objects such as density perturbations are entirely
absent.
The corresponding symmetric tensor is constructed out of derivatives 
as
\begin{equation}
Q_{ij}^\vepm = - {1 \over 2k}
	(\nabla_i Q_j^\vepm + \nabla_j Q_i^\vepm )\, .
\end{equation}
A convenient representation is 
\begin{equation}
{Q}_i^\vepm = - {i \over \sqrt{2}} 
		(\hat{e}_1 \pm i \hat{e}_2)_i \exp(i\vec{k}
	\cdot \vec{x}) \, .
\end{equation}
Notice that $n^i Q_i^\vepm = G_1^{\pm 1}$ and
$n^i n^j Q_{ij}^\vepm \propto G_2^{\pm 1}$.  
Thus vector perturbations stimulate
the $m=\pm 1$ modes in the radiation.  

\subsubsection{Tensor Perturbations}

Tensor perturbations are represented by Laplacian eigenfunctions
\begin{equation}
\nabla^2 Q_{ij}^\tepm = -k^2 Q_{ij}^\tepm ,
\label{eqn:tensoreigen}
\end{equation}
which satisfy a transverse-traceless condition
\begin{equation}
\delta^{ij} Q^\tepm_{ij} = \nabla^i Q^\tepm_{ij}{} = 0\, ,
\end{equation}
that forbids the construction of scalar and vector objects
such as density and velocity fields. The modes take on
an explicit representation of
\begin{equation}
{Q}_{ij}^\tepm = - \sqrt{3 \over 8} 
	      (\hat{e}_1 \pm i \hat{e}_2 )_i \otimes 
	      (\hat{e}_1 \pm i \hat{e}_2 )_j 
		\exp(i \vec{k} \cdot \vec{x})\, .
\end{equation}
Notice that  $n^i n^j Q_{ij}^\tepm = G_2^{\pm 2}$ 
and thus tensors stimulate the
$m=\pm 2$ modes in the radiation.

In the following sections,  
we often only explicitly show
the positive $m$ value with the understanding that its opposite
takes on the same form except where otherwise noted (i.e.~in
the $B$-type polarization where a sign reversal occurs).  

\section{Perturbation Evolution} \label{sec:pertevol}

We discuss here the evolution of perturbations in the normal
modes of \S \ref{sec:normal}.  We first review the decomposition
of perturbations in the metric and stress-energy tensor into
scalar, vector and tensor types (\S \ref{sec:perturbations}).  We
further divide the stress-energy tensor into fluid contributions,
applicable to the usual particle species,
and seed perturbations, applicable to cosmological defect models.  
We then employ the 
techniques developed in \S \ref{sec:normal}   
to obtain a new, simpler derivation and form of the 
radiation transport
of the CMB under Thomson scattering, including polarization 
(\S \ref{sec:transport}), 
than that obtained first by \cite{Cha}.
The complete evolution equations, given in
\S \ref{sec:evolution}, are again substantially simpler in form
than those of prior works where they overlap
\cite{SelZal,KamKosSte,Pol,BonEfs}
and treats the case of vector perturbations. Finally
in \S \ref{sec:integral}, we derive the formal integral solutions
through the use of the radial functions of \S \ref{sec:radial}
and discuss their geometric interpretation.  These solutions
encapsulate many of the important results.  

\subsection{Perturbations}\label{sec:perturbations}

\subsubsection{Metric Tensor} \label{sec:metric}

The ultimate source of CMB anisotropies is the gravitational 
redshift induced by the metric fluctuation $h_{\mu\nu}$
\begin{equation}
g_{\mu\nu} = a^2(\eta_{\mu\nu} + h_{\mu\nu}) \, ,
\label{eqn:genmetric}
\end{equation}
where the zeroth component represents conformal time 
$d\eta = dt/a$ and, in the flat universe considered here,
 $\eta_{\mu\nu}$ is the Minkowski metric.
The metric perturbation can be further broken up into the
normal modes of
scalar, vector and tensor types as in \S \ref{sec:classification}.
Scalar and vector modes
exhibit gauge freedom which is fixed by an explicit choice of
the coordinates that relate the perturbation to the background.
For the scalars, we choose the Newtonian gauge (see e.g. 
\cite{KodSas,Bar})
\begin{equation}
h_{00} = 2\Psi Q^\se , \qquad h_{ij} = 2\Phi Q^\se \delta_{ij}\, ,
\label{eqn:smetric}
\end{equation}
where the metric is shear free.  For the vectors, we choose
\begin{equation}
h_{0i} = -V Q_i^\ve ,
\label{eqn:vmetric}
\end{equation}
and the tensors
\begin{equation}
h_{ij} = 2 H Q_{ij}^\te .
\label{eqn:tmetric}
\end{equation}
Note that tensor fluctuations do not exhibit gauge freedom of
this type.

\subsubsection{Stress Energy Tensor}

The stress energy tensor can be broken up into fluid ($f$) 
contributions
and seed ($s$) contributions (see e.g.~\cite{Dur}).  
The latter is distinguished by the fact
that the net effect can be viewed as a perturbation to the
background.  Specifically $T_{\mu\nu} = \bar T_{\mu\nu} + 
\delta T_{\mu\nu}$ where 
$\bar T^0_{\hphantom{0}0} = -\rho_f$,
$\bar T^0_{\hphantom{0}i} = \bar T_0^{\hphantom{i}i} =0$
and $\bar T^i_{\hphantom{i}j} = p_f \delta^i {}_j$
is given by the fluid alone.   The fluctuations can be 
decomposed into the normal modes of \S \ref{sec:classification} as
\begin{equation}
\begin{array}{lcl}
\delta T^0_{\hphantom{0}0} &=&
        - [\rho_f \delta_f + \rho_s] \, Q^\se, \vertsp\\
\delta T^0_{\hphantom{0}i} &=& [(\rho_f + p_f) v_f^\se
	+ v_s^\se]
        \, Q_i^\se  ,
        \vertsp\\
\delta T_0^{\hphantom{i}i} &=& -[(\rho_f + p_f)v_f^\se + v_s^\se] \,
        Q^\se{}^i,
        \vertsp\\
\delta T^i_{\hphantom{i}j} &=& [\delta p_f + p_s
        \delta^i_{\hphantom{i}j}] Q^\se + [ p_f\pi_f + p_s]
        Q^\se{}^i_{\hphantom{i}j} \vertsp \, .
\end{array}
\label{eqn:sstress}
\end{equation}
for the scalar components,
\begin{equation}
\begin{array}{rcl}
 \delta T^i_{\hphantom{0}0} & = & -[(\rho_f + p_f)
        v_f^\ve + \rho_s] \, Q^{\ve\, i} , \vertsp\\
 \delta T^0_{\hphantom{0}i}  & = & [(\rho_f+p_f)
        (v_f^\ve - V) + v_s^\ve] \, Q^\ve_i , \vertsp\\
 \delta T^i_{\hphantom{0}j}  & = & [p_f\pi_f^\ve + \pi_s^\ve]
        \, Q^\ve{}^i_{\hphantom{i}j}\vertsp\,  ,
\label{eqn:vstress}
\end{array}
\end{equation}
for the vector components, and
\begin{equation}
\delta T^i_{\hphantom{0}j} =   [p_f\pi_f^{\te} + \pi_s^\te] \,
        Q^\te {}^i_{\hphantom{i}j} \, ,
\label{eqn:tstress}
\end{equation}
for the tensor components.

\subsection{Radiation Transport} \label{sec:transport}

\subsubsection{Stokes Parameters}  

The Boltzmann equation for the CMB describes the transport
of the photons under Thomson scattering by the electrons.  The
radiation is described by the intensity matrix: the time
average of the electric field tensor $E_i^* E_j$ over a time
long compared to the frequency of the light or equivalently as the 
components of the photon density matrix (see \cite{Kos} for
reviews).  For radiation
propagating radially $\vec{E} \perp \hat{e}_r$, so that
the intensity matrix exists on the $\hat{e}_\theta 
\otimes \hat{e}_\phi$ subspace.  The matrix can further
be decomposed in terms of the $2 \times 2$ Pauli matrices $\sigma_i$ 
and the unit matrix ${\bf 1}$ on this subspace.  

For our purposes, it is convenient to describe the polarization
in temperature fluctuation units rather than intensity,
where the analogous matrix becomes,
\begin{equation}
{\bf T} = \Theta {\bf 1} + Q \sigma_3 + U \sigma_1
                + V \sigma_2.
\label{eqn:tempmatrix}
\end{equation}
$\Theta = {\rm Tr}({\bf T} {\bf 1})/2 = \Delta T/T$ is the 
temperature perturbation summed
over polarization states.  Since $Q = {\rm Tr}({\bf T}\sigma_3)/2$,
it is the difference in temperature fluctuations polarized in
the $\hat{e}_\theta$ and $\hat{e}_\phi$ directions.  
Similarly $U={\rm Tr}({\bf T} \sigma_1)/2$ is the difference along
axes rotated by 45$^\circ$, $(\hat{e}_\theta 
\pm \hat{e}_\phi)/\sqrt{2}$,
and $V={\rm Tr}({\bf T} \sigma_2)/2$ that between $(\hat{e}_\theta 
\pm i \hat{e}_\phi)/\sqrt{2}$.
$Q$ and $U$ thus represent linearly polarized
light in the north/south--east/west and 
northeast/southwest--northwest/southeast directions on the sphere
respectively. $V$ represents circularly polarized light 
(in this section only,
not to be confused with vector metric perturbations). 

Under a counterclockwise rotation of the axes
through an angle $\psi$ the intensity ${\bf T}$ transforms as
${\bf T}'={\bf R}{\bf T}{\bf R}^{-1}$.  
$\Theta$ and $V$ remain distinct
while $Q$ and $U$ transform into one another.  Since the
Pauli matrices transform as $\sigma_3' \pm i \sigma_1'
= e^{\mp 2i \psi} (\sigma_3 \pm i \sigma_1)$ a more convenient
description is
\begin{equation}
{\bf T} = \Theta {\bf 1} + V \sigma_2 +
	(Q + iU) {\bf M}_+ 
+	(Q - iU) {\bf M}_- \, ,
\label{eqn:polqiu}
\end{equation}
where recall that ${\bf M}_{\pm} = (\sigma_3 \mp i \sigma_1)/2$
(see Eqn.~\ref{eqn:matrixbasis}),
so that $Q \pm i U$ transforms into itself under rotation.
Thus Eqn.~(\ref{eqn:spinbasis}) implies that $Q \pm i U$ should
be decomposed into $s= \pm 2$ spin harmonics \cite{SelZal,KamKosSte}.

Since circular polarization cannot be generated by 
Thomson scattering alone, we shall hereafter ignore $V$. It is then 
convenient to reexpress the matrix as a vector 
\begin{equation}
\vec{T} = (\Theta, Q+iU, Q-iU) \, .
\label{eqn:Trep}
\end{equation}
The Boltzmann equation describes the evolution of the vector
$\vec{T}$ under the Thomson collisional term $C[\vec{T}]$
and gravitational redshifts in a perturbed metric $G[h_{\mu\nu}]$
\begin{equation}
{d \over d\eta} \vec{T} (\eta,\vec{x},\hat{n}) \equiv
{\partial \over \partial \eta} \vec{T} + n^i \nabla_i \vec{T} 
=  \vec{C} [{\vec T}]
+ \vec{G} [h_{\mu\nu}] \, ,
\label{eqn:boltzmannimplicit}
\end{equation}
where we have used the fact that $\dot x_i = n_i$ and that
in a flat universe photons propagate in straight lines $\dot n=0$.
We shall now evaluate the Thomson scattering and gravitational
redshift terms.

\subsubsection{Scattering Matrix} \label{sec:scattering}

The
calculation of Thomson scattering including polarization was
first performed by Chandrasekhar \cite{Cha};
here we show a much simpler derivation employing the spin
harmonics.
The Thomson differential scattering cross section 
depends on angle as 
$|\hat{\epsilon}' \cdot \hat{\epsilon}|^2$ where $\hat{\epsilon}'$ 
and $\hat{\epsilon}$ are the incoming and outgoing polarization
vectors respectively in the electron rest frame.  
Radiation polarized perpendicular to the
scattering plane scatters isotropically, 
while that in the scattering plane
picks up a factor of $\cos^2\beta$ where $\beta$ is the scattering
angle.  
If the radiation has different intensities or temperatures
at right angles, the radiation scattered into a given angle
will be linearly polarized.

Now let us evaluate the scattering term explicitly.  
The angular dependence of the scattering gives
\begin{equation}
\left( \begin{array}{c}
\Theta_{\parallel} \\
\Theta_{\perp} \\ 
U     \end{array}\right)'
= 
\left( \begin{array}{ccc}
\cos^2\beta & 0 & 0 \\
0 & 1 & 0 \\
0 & 0 & \cos\beta \end{array} \right)
\left( \begin{array}{c}
\Theta_{\parallel} \\
\Theta_{\perp} \\
U \\ \end{array} \right) \, ,
\end{equation}
where the  $U$ transformation follows from its definition 
in terms of the 
difference in intensities polarized $\pm 45^\circ$ from the
scattering plane.
With the relations $\Theta=\Theta_\parallel + \Theta_\perp$ and
$Q \pm iU = \Theta_\parallel - \Theta_\perp \pm iU$, 
the angular dependence
in the $\vec{T}$ representation of Eqn.~(\ref{eqn:Trep})
becomes,\footnote{Chandrasekhar employs a different
sign convention for $U \rightarrow -U$.}  
\begin{equation}
\vec{T}'= {\bf S}\vec{T} =
{3 \over 4}\left(
\begin{array}{ccc}
\cos^2\beta +1  \quad & -{1 \over 2}\sin^2\beta  \quad 
		       & -{1 \over 2}\sin^2\beta \vertsp\\
-{1 \over 2}\sin^2  \beta
	         \quad & {1 \over 2}(\cos\beta + 1)^2 \quad 
		       & {1 \over 2}(\cos\beta - 1)^2 \vertsp\\
-{1 \over 2}\sin^2 \beta
	         \quad & {1 \over 2}(\cos\beta - 1)^2 \quad 
		       & {1 \over 2}(\cos\beta + 1)^2 \vertsp\\
\end{array}
\right) \vec{T} \, ,
\label{ref:scatmatrix}
\end{equation}
where the overall normalization is fixed by photon conservation in
the scattering.  To relate these scattering frame quantities to those
in the frame defined by $\hat{k} = \hat{e}_3$, we must first perform
a rotation from the $\hat{k}$ frame to the scattering frame.
The geometry is displayed in Fig.~\ref{fig:scatgeom}, where
the angle $\alpha$ separates the scattering plane from
the meridian plane at $(\theta',\phi')$ 
spanned by $\hat{e}_r$ and $\hat{e}_\theta$.  
After scattering, we rotate by
the angle between the scattering plane and the meridian plane
at $(\theta,\phi)$ to return to the $\hat{k}$ frame.
Eqn.~(\ref{eqn:polqiu}) tells us these rotations
transform  $\vec{T}$ as ${\bf R}(\psi)\vec{T} = {\rm diag}
(1, e^{2i\psi}, e^{-2i\psi}) \vec{T}$.  
The net result is thus expressed as 
\begin{equation}
{\bf R}(\gamma){\bf S}(\beta){\bf R}(-\alpha) =
{1 \over 2}{\sqrt{4\pi \over 5}}\left(
\begin{array}{ccc}
Y_2^0(\beta,\alpha)  + 2 \sqrt{5} Y_0^0(\beta,\alpha) \quad 
	& -\sqrt{3 \over 2} Y_2^{-2}(\beta,\alpha) \quad 
        & -\sqrt{3 \over 2} Y_2^2(\beta,\alpha) \vertsp\\
-\sqrt{6} \Spy{2}{2}{0}(\beta,\alpha)e^{-2i\gamma} \quad 
	& 3 \Spy{2}{2}{-2}(\beta,\alpha)e^{-2i\gamma} \quad
	& 3 \Spy{2}{2}{2}(\beta,\alpha)e^{-2i\gamma} \vertsp\\
-\sqrt{6} \Spy{-2}{2}{0}(\beta,\alpha)e^{2i\gamma} \quad 
	& 3 \Spy{-2}{2}{-2}(\beta,\alpha)e^{2i\gamma} \quad
	& 3 \Spy{-2}{2}{2}(\beta,\alpha)e^{2i\gamma} \vertsp
\end{array}
\right) \, , 
\label{eqn:RSR}
\end{equation}
where we have employed the explict spin-2, $\ell=2$ forms in Tab.~1.
Integrating over incoming angles, we obtain the collision term
in the electron rest frame
\begin{equation}
\vec{C}[{\vec {T}}]_{\rm rest} 
= -\dot\tau \vec{T}(\Omega) + \dot\tau \int {d\Omega' \over 4\pi}
 {\bf R}(\gamma){\bf S}(\beta){\bf R}(-\alpha) \vec{T}(\Omega')\, ,
\label{eqn:collisionimplicit}
\end{equation}
where the two terms on the rhs 
account for scattering out of and into
a given angle respectively.   Here the differential optical
depth $\dot \tau = n_e \sigma_T a$ sets the collision rate
in conformal time with $n_e$ as the free electron density
and $\sigma_T$ as the Thomson cross section.  

The transformation from the electron rest frame into the
background frame yields a Doppler shift 
$\hat{n} \cdot \vec{v}_B$ in the temperature of the scattered radiation.
With the help of the generalized addition relation for the harmonics
Eqn.~(\ref{eqn:composition}), the full collision term can be written
as
\begin{equation}
\vec{C}[{\vec {T}}] = -\dot\tau {\vec I}(\Omega) + 
        {1 \over 10}\dot\tau \int d\Omega'
	\sum_{m=-2}^2 {\bf P}^{(m)}(\Omega,\Omega') \vec{T}(\Omega')\, .
\label{eqn:fullcollision}
\end{equation}
The vector $\vec{I}$ describes
the isotropization of distribution in the 
electron rest frame and is given by
\begin{equation}
\vec{I}(\Omega) = \vec{T}(\Omega)
 - \left( \int {d\Omega'\over 4\pi} 
	\Theta' + \hat{n} \cdot \vec{v}_B \, , 0 , \, 0 \right) .
\end{equation}
The matrix ${\bf P}^{(m)}$ encapsulates the anisotropic nature
of Thomson scattering and shows that as expected
polarization is generated through quadrupole anisotropies 
in the temperature and vice versa
\begin{eqnarray}
{\bf P}^{(m)} =
\left(
\begin{array}{ccc}
Y_2^{m}{}'\, Y_2^m   \quad &
          - \sqrt{3 \over 2} \Spy{2}{2}{m}{}'\, Y_2^m 
        \quad & - \sqrt{3 \over 2} 
	  \Spy{-2}{2}{m}{}'\, Y_2^m 
	\vertsp\\
- \sqrt{6} Y_2^{m}{}' \Spy{2}{2}{m} \quad &
3 \Spy{2}{2}{m}{}'\Spy{2}{2}{m} \quad &
          3 \Spy{-2}{2}{m}{}'\Spy{2}{2}{m}{} \quad 
		\vertsp\\
- \sqrt{6} Y_2^{m}{}' \Spy{-2}{2}{m}{} \quad &
          3 \Spy{2}{2}{m}{}' \Spy{-2}{2}{m} \quad &
          3 \Spy{-2}{2}{m}{}' \Spy{-2}{2}{m}
		\vertsp\\
\end{array}
\right),
\label{eqn:scatmatrix}
\end{eqnarray}
where $Y_\ell^m {}'\equiv Y_\ell^{m*} (\Omega')$ 
and   $\Spy{s}{\ell}{m} {}' \equiv \Spy{s}{\ell}{m*}(\Omega')$
and the unprimed harmonics are with respect to $\Omega$.
These $m =0,\pm 1, \pm 2$ components correspond to the 
scalar, vector and tensor scattering terms as discussed in 
\S \ref{sec:classification} and \ref{sec:evolution}.  

\subsubsection{Gravitational Redshift}

In a perturbed metric, gravitational
interactions alter the temperature perturbation $\Theta$.  
The redshift properties may be formally derived by employing the
equation of motion for the photon energy $p \equiv -u^\mu p_\mu$
where $u^\mu$ is the 4-velocity of an observer at rest
in the background frame and $p^\mu$ is the photon 4-momentum. 
The Euler-Lagrange
equations of motion for the photon and the requirement
that $|u^2|=1$ result in 
\begin{equation}
{\dot{p}\over p} = -\dot{a \over a} - {1 \over 2}
         {n}^i {n}^j \dot h_{ij}
- {n}^i \dot h_{0i} 
        - {1 \over 2} n^i \nabla_i h_{00} \, ,
\label{eqn:SWgeneral}
\end{equation}
which differs from \cite{SacWol,AbbSch} since we take $\hat{n}$
to be the photon propagation direction rather than the viewing
direction of the observer.
The first term is the cosmological reshift due to the expansion
of the spatial metric; it does not affect temperature perturbations
$\delta T/T$.  The second term has a similar origin and is due
to stretching of the spatial metric.  The third and fourth term
are the frame dragging and time dilation effects.

Since gravitational redshift affects the different polarization
states alike, 
\begin{equation}
\vec{G}[h_{\mu\nu}] = \left({1 \over 2}
         {n}^i {n}^j \dot h_{ij}
+ {n}^i \dot h_{0i}
        + {1 \over 2} n^i \nabla_i h_{00} \,,  0 \, , 0 \right) ,
\label{eqn:gravred}
\end{equation}
in the $\vec{T}$ basis.
We now explicitly evaluate the Boltzmann equation for scalar,
vector, and tensor metric fluctuations of
Eqns.~(\ref{eqn:smetric})-(\ref{eqn:tmetric}).

\subsection{Evolution Equations} \label{sec:evolution}

In this section, we derive the  complete set of evolution
equations
for the temperature and polarization distribution 
in the scalar, vector, tensor decomposition of metric fluctuations.
Though the scalar and tensor fluid results can be found elsewhere
in the literature in a different form
(see e.g.~\cite{BonEfs,Pol}), the total
angular momentum representation substantially simplifies the 
form and aids in the interpretation of the results.  The
vector derivation is new to this work.  

\subsubsection{Angular Moments and Power}

The temperature and polarization fluctuations
are expanded into the normal modes defined 
in \S \ref{sec:radial},\footnote{Our conventions differ
from \cite{SelZal} as $(2\ell+1)\Delta_{T\ell}^{(S,T)} =
4\Theta_\ell^{(0,2)}/(2\pi)^{3/2}$ and similarly
for $\Delta_{E,B\ell}^{(S,T)}$ with $\Theta_\ell^{(0,2)} \rightarrow
-E_\ell^{(0,2)}, -B_\ell^{(0,2)}$ and so $C_{C\ell}^{(S,T)} = 
- C_\ell^{\Theta E(0,2)}$ but with other power spectra the same.}
\begin{equation}
\begin{array}{rcl}
\Theta(\eta,\vec{x},\vec{n}) &=& \displaystyle{ 
\int {d^3 k \over (2\pi)^3} }
	\sum_{\ell}
\sum_{m=-2}^2 \Theta_\ell^{(m)} G_\ell^m \, , \\
 (Q \pm i U)(\eta,\vec{x},\vec{n}) &=& \displaystyle{\int {d^3k 
\over (2\pi)^3}}
	\sum_{\ell} \sum_{m=-2}^2
	(E_\ell^{(m)} \pm i B_\ell^{(m)}) \, {}_{\pm 2} G_\ell^m .
\end{array}
\label{eqn:decomposition}
\end{equation}
A comparison with Eqn.~(\ref{eqn:parityeigenstates}) and 
(\ref{eqn:polqiu})  shows
that $E_\ell^{(m)}$ and $B_\ell^{(m)}$ represent 
polarization with electric $(-1)^\ell$ and magnetic $(-1)^{\ell+1}$
type parities
respectively \cite{SelZal,KamKosSte}.  Because the temperature
$\Theta^{(m)}_\ell$ has electric type parity, only $E_\ell^{(m)}$ 
couples
directly to the temperature in the scattering sources.
Note that $B_\ell^{(m)}$ and $E_\ell^{(m)}$ represent polarizations
with $Q$ and $U$ interchanged and thus represent polarization
patterns rotated by $45^\circ$.  A simple example is given by
the $m=0$ modes.  In the $\hat{k}$-frame,
$E_\ell^{(0)}$ represents a pure $Q$,
or north/south--east/west, polarization field 
whose amplitude depends on 
$\theta$, e.g.~$\sin^2\theta$ for $\ell=2$.  $B_\ell^{(0)}$
represents a pure $U$, or northwest/southeast--northeast/southwest, 
polarization with the
same dependence.

The power spectra of temperature and polarization anisotropies
today are defined as, e.g.~$C_\ell^{\Theta\Theta} \equiv 
\left< | a_{\ell m} |^2 \right>$ for $\Theta = \sum a_{\ell m}
Y_\ell^m$ with the average being over
the ($2\ell+1$) $m$-values.  Recalling the normalization of the
mode functions from Eqn.~(\ref{eqn:temperaturebasis}) and 
(\ref{eqn:polarizationbasis}), we obtain
\begin{equation}
(2\ell+1)^2 C_\ell^{X\widetilde X} 
		   = {2 \over \pi}
	\int {dk \over k}
	\sum_{m=-2}^2
        k^3 X_\ell^{(m)*}(\eta_0,k) \widetilde X_\ell^{(m)}(\eta_0,k)\, , 
\label{eqn:cl}
\end{equation}
where $X$ takes on the values $\Theta$, $E$ and $B$.
There is no cross correlation $C_\ell^{\Theta B}$ or
$C_\ell^{E B}$ due to parity [see Eqns.~(\ref{eqn:parity}) and
(\ref{eqn:ebparity})].  
We also employ the notation $C_\ell^{X\widetilde X(m)}$ for the $m$ contributions
individually.
Note that $B_\ell^{(0)}=0$ here due to 
azimuthal symmetry in the transport problem so that $C_\ell^{BB(0)}
=0$.

As we shall now show, the $m=0,\pm 1,\pm 2$ modes are stimulated by
scalar, vector and tensor perturbations in the metric. The 
orthogonality of the spherical harmonics assures us that these modes
are independent, and we now discuss the contributions separately.

\subsubsection{Free Streaming}

As the radiation free streams, gradients in the distribution
produce anisotropies.  For example, as photons from different 
temperature regions intersect on their trajectories, the temperature
difference is reflected in the angular distribution.  
This effect is represented in the Boltzmann equation 
(\ref{eqn:boltzmannimplicit}) gradient term,
\begin{equation}
\hat{n} \cdot \vec{\nabla} \rightarrow
i\hat{n} \cdot \vec{k} = i\sqrt{4\pi \over 3}k  Y_1^0 \, .
\end{equation}
which multiplies the intrinsic angular dependence of the
temperature and polarization distributions, 
$Y_\ell^m$ and $\Spy{\pm 2}{\ell}{m}$ respectively,
from the expansion Eqn.~(\ref{eqn:decomposition}) and 
the angular basis of Eqns.~(\ref{eqn:temperaturebasis}) 
and (\ref{eqn:polarizationbasis}).  Free streaming
thus involves the 
Clebsch-Gordan relation of Eqn.~(\ref{eqn:ClebschGordan})
\begin{eqnarray}
\sqrt{4 \pi \over 3} Y_1^0 (\Spy{s}{\ell}{m}) 
   & = & {\kap{s}{\ell}{m} \over \sqrt{(2\ell+1)(2\ell-1)}}
         \left(\Spy{s}{\ell-1}{m}\right)
- {m s \over \ell (\ell +1)} \left({}_s Y_{\ell}^m\right)
+ {\kap{s}{\ell+1}{m} \over \sqrt{(2\ell+1)(2\ell+3)}}
	\left(\Spy{s}{\ell+1}{m}\right)
\label{eqn:streamingcg}
\end{eqnarray}
which couples the $\ell$ to $\ell\pm 1$ moments of the distribution.
Here the coupling coefficient is
\begin{equation}
\kap{s}{\ell}{m} = \sqrt{(\ell^2-m^2)(\ell^2-s^2)/\ell^2}\, .
\label{eqn:kapfactor}
\end{equation}
As we shall now see, the result of this streaming effect is
an infinite hierarchy of coupled $\ell$-moments 
that passes power from sources at low multipoles up 
the $\ell$-chain as time progresses.

\subsubsection{Boltzmann Equations}

The explicit form of the
Boltzmann equations for the temperature and polarization
follows directly from the Clebsch-Gordan relation of 
Eqn.~(\ref{eqn:streamingcg}).
For the temperature ($s=0$),
\begin{equation}
\dot \Theta_\ell^{(m)}
= k\Bigg[ { \kap{0}{\ell}{m}
                \over (2\ell-1)}
        \Theta_{\ell-1}^{(m)}
             -{ \kap{0}{\ell+1}{m}
                \over (2\ell+3)}
        \Theta_{\ell+1}^{(m)} \Bigg]
        - \dot\tau \Theta_\ell^{(m)} + S_\ell^{(m)},  \qquad (\ell \ge m).
\label{eqn:boltz}
\end{equation}
The term in the square brackets is the free streaming effect
that couples the $\ell$-modes and tells us that in the absence of
scattering power is transferred down the hierarchy when $k\eta
\simgt 1$.  This transferral merely represents geometrical
projection of fluctuations on the scale corresponding to $k$ at 
distance $\eta$ which subtends an angle given by $\ell \sim k\eta$. 
The main effect of scattering 
comes through the $\dot\tau \Theta_\ell^{(m)}$ term and implies
an exponential suppression of anisotropies with optical depth in
the absence of sources.  The source $S_\ell^{(m)}$ accounts for
the gravitational and residual scattering effects,
\begin{equation}
\begin{array}{lll}
S_0^{(0)} = \dot\tau \Theta_0^\se - \dot \Phi \, , 	\qquad &
S_1^{(0)} = \dot\tau v_B^\se + k\Psi \, , 		\qquad &
S_2^{(0)} = \dot\tau P^\se \, , \vertsp\\
						\qquad &
S_1^{(1)} = \dot\tau v_B^\ve + \dot V \, ,		\qquad &
S_2^{(1)} = \dot\tau P^\ve \, , \vertsp\\
						\qquad &
						\qquad &
S_2^{(2)} = \dot\tau P^\te - \dot H	 \vertsp \, .
\end{array}
\label{eqn:tempsources}
\end{equation}
The presence of $\Theta_0^\se$ represents the fact that
an isotropic temperature fluctuation is not destroyed by
scattering.  The Doppler effect enters the dipole $(\ell=1)$
equation through the baryon velocity
$v_B^{(m)}$ term.  Finally the anisotropic
nature of Compton scattering is expressed through
\begin{equation}
P^{(m)} = {1 \over 10} \left[ \Theta_2^{(m)}  -
\sqrt{6} E_2^{(m)} \right], 
\label{eqn:polsource}
\end{equation}
and involves the quadrupole moments of the temperature and 
$E$-polarization
distribution only.

The polarization evolution follows a similar pattern for 
$\ell \ge 2$, $m\ge 0$ from Eqn.~(\ref{eqn:streamingcg}) with
$s=\pm 2$,\footnote{The expressions above were 
all derived assuming a flat spatial
geometry.  In this formalism, including the effects of spatial
curvature is straightforward: the $\ell\pm 1$ terms in the 
hierarchy are multiplied by factors of
$[1-(\ell^2 - m-1)K/k^2]^{1/2}$
 \cite{Tom,AbbSch},
where the curvature is $K=-H_0^2(1-\Omega_{\rm tot})$.  These
factors account for geodesic deviation and alter the transfer of
power through the hierarchy.  A full treatment of such effects
will be provided in \cite{HSWZ}.}

\begin{eqnarray}
\dot E_\ell^{(m)} &=& 
	k \Bigg[ 
	{\kap{2}{\ell}{m} \over (2\ell-1)}
E_{\ell-1}^{(m)} - {2m \over \ell (\ell + 1)} B_\ell^{(m)} 
- {\kap{2}{\ell+1}{m} \over 
	(2 \ell + 3)}
        E_{\ell + 1}^{(m)}
\Bigg] - \dot\tau [E_\ell^{(m)} + \sqrt{6} P^{(m)} \delta_{\ell,2}]\, , \\ 
\dot B_\ell^{(m)} &=& 
	k \Bigg[ 
        {\kap{2}{\ell}{m} \over (2\ell-1)}
B_{\ell-1}^{(m)} + {2m \over \ell (\ell + 1)} E_\ell^{(m)} 
- {\kap{2}{\ell+1}{m} \over
        (2 \ell + 3)}
        B_{\ell + 1}^{(m)}
\Bigg] - \dot\tau B_\ell^{(m)} .  
\label{eqn:epol}
\end{eqnarray}
Notice that the source of polarization $P^{(m)}$
enters only in the $E$-mode quadrupole 
due to the opposite parity of $\Theta_2$
and $B_2$.  However, as discussed in \S \ref{sec:radial}, free
streaming or projection couples the two parities except for
the $m=0$ scalars.  Thus $B_\ell^{(0)} =0$ by geometry 
regardless of the source.
It is unnecessary to solve separately for the $m=-|m|$ relations
since they satisfy the same equations and solutions with
$B_\ell^{(-|m|)} = -B_\ell^{(|m|)}$ and all other quantities 
equal.

To complete these equations, we need to express the evolution of
the metric sources $(\Phi,\Psi,V,H)$.  It is to this subject we
now turn.

\subsubsection{Scalar Einstein Equations}

The Einstein equations
$G_{\mu\nu} = 8\pi G T_{\mu\nu}$ express the metric evolution 
in terms of the matter sources.
With the
form of the scalar metric and stress energy tensor given
in Eqns.~(\ref{eqn:smetric}) and (\ref{eqn:sstress}), the ``Poisson''
equations become
\begin{equation}
\begin{array}{rcl}
k^2 \Phi &=& 4\pi G a^2 \left[ (\rho_f \delta_f + \rho_s)
        + 3
\displaystyle{\dot a \over a}[ (\rho_f + p_f)v_f^\se + v_s^\se]
/k \right] , \\
k^2 (\Psi + \Phi) &=& -8\pi G a^2 \left( p_f \pi_f^\se + 
	\pi_s^\se \right),
\vertsp
\end{array}
\label{eqn:Poisson}
\end{equation}
where the corresponding matter evolution is
given by covariant conservation of the stress energy tensor
$T_{\mu\nu}$,
\begin{eqnarray}
\dot \delta_f 
& = & -(1+w_f)(kv_f^\se + 3\dot\Phi) - 3{\dot a \over a} \delta w_f 
	\, , \nonumber\\
{d \over d\eta} \left[ (1 + w_f) v_f^\se \right] 
& = & (1+w_f) \left[ k\Psi - {\dot a  \over a} (1- 3w_f) v_f^\se\right]
	+ w_f k(\delta p_f/p_f - {2 \over 3} \pi_f ) \, ,
\label{eqn:sfluideqn}
\end{eqnarray}
for the fluid part, where $w_f = p_f/\rho_f$.  These equations 
express energy 
and momentum density conservation respectively.
They remain true for each fluid individually in the absence
of momentum exchange.  Note
that for the photons
$\delta_\gamma = 4\Theta^\se_0$, $v_\gamma^\se = \Theta_1^\se$
and $\pi_\gamma^\se = {12 \over 5}\Theta_2^\se$.  Massless
neutrinos obey Eqn.~(\ref{eqn:boltz}) without the Thomson
coupling term.   

Momentum exchange
between the baryons and photons due to Thomson scattering 
follows by noting that for a given velocity perturbation
the momentum density ratio between the two fluids is 
\begin{equation}
R \equiv {\rho_B+p_B \over \rho_\gamma + p_\gamma} \approx 
{3\rho_B \over 4\rho_\gamma} \, .
\label{eqn:Rdef}
\end{equation} 
A comparison with photon Euler equation~(\ref{eqn:boltz}) (with
$\ell=1$, $m=0$) gives the
baryon equations as
\begin{eqnarray}
\dot \delta_B & = & - kv_B^\se - 3\dot\Phi \, ,\nonumber\\
\dot v_B^\se &=& -{\dot a \over a} v_B^\se + k\Psi + {\dot\tau \over R}
	(\Theta_1^\se - v_B^\se) \, .
\label{eqn:sbaryon}
\end{eqnarray}
For a seed source, the conservation equations become
\begin{eqnarray}
\dot \rho_s &=& 
	-3{\dot a \over a} (\rho_s + p_s) - kv_s^\se , \nonumber\\  
\dot v_s^\se &=&
	-4{\dot a \over a} v_s^\se + k(p_s -{2 \over 3}\pi_s^\se)\, ,
\end{eqnarray}
since the metric fluctuations produce higher order terms.

\subsubsection{Vector Einstein Equations}

The vector metric source evolution is similarly constructed from
a ``Poisson'' equation 
\begin{equation}
\dot V + 2 {\dot a \over a} V =
        -8\pi G a^2 (p_f \pi^\ve_f + \pi_s^\ve)/k \, ,
\end{equation}
and the momentum conservation equation for 
the stress-energy tensor or Euler equation 
\begin{eqnarray}
 \dot v_f^\ve &= &\dot V  - (1 - 3c_f^2 )
{\dot a \over a} (v^\ve_f - V)  - {1 \over 2 } k
{w_f \over 1+w_f} \pi^\ve_f , \nonumber\\
\dot v_s^\ve &=& - 4{\dot a \over a} v_s^\ve  -
        {1 \over 2}k\pi_s^\ve, 
\label{eqn:vectoreulergen}
\end{eqnarray}
where recall $c_f^2 = \dot p_f / \dot \rho_f$ is the sound speed.
Again, the first of these equations remains true for each fluid
individually save for momentum exchange terms. 
For the photons $v_\gamma^\ve = \Theta_1^\ve$ and
$\pi_\gamma^\ve = {8 \over 5}\sqrt{3}\Theta_2^\ve$.
Thus with the photon Euler equation 
(\ref{eqn:boltz}) (with $\ell=1$, $m=1$), 
the full  baryon equation becomes
\begin{equation}
 \dot v^\ve_B = \dot V  -  {\dot a \over a}
 (v^\ve_B - V) + {\dot\tau \over R} (\Theta_1^\ve - v^\ve_B) \, ,
\label{eqn:vbaryon}
\end{equation}
see Eqn.~(\ref{eqn:sbaryon}) for details.

\subsubsection{Tensor Einstein Equations}

The Einstein equations tell us that the tensor metric source is governed
by
\begin{equation}
\ddot H + 2{\dot a \over a} \dot H +
k^2 H = 8\pi G a^2 [
p_f^{\vphantom{\te}} \pi^{\te}_f + \pi^{\te}_s]\, ,
\end{equation}
where note that the photon contribution is 
$\pi^\te_\gamma = {8 \over 5}\Theta_2^\te$.

\subsection{Integral Solutions} \label{sec:integral}

The Boltzmann equations have formal integral solutions that 
are simple to write down by considering the properties of 
source projection from \S \ref{sec:radial}.  The
hierarchy equations for the temperature 
distribution Eqn.~(\ref{eqn:boltz}) merely express the 
projection of the various
plane wave temperature sources $S_\ell^{(m)} G_\ell^m$ on the sky
today (see Eqn.~(\ref{eqn:tempsources})).  
From the angular decomposition of $G_\ell^m$ in 
Eqn.~(\ref{eqn:radialtemp}), the integral solution immediately follows
\begin{equation}
{\Theta_\ell^{(m)}(\eta_0,k) \over 2\ell + 1}\, 
  =  
\int_0^{\eta_0} d\eta \, e^{-\tau} \sum_{\ell'} \,
  S_{\ell'}^{(m)}(\eta) \, j_\ell^{(\ell'm)}(k(\eta_0-\eta)) \, .
\label{eqn:inttemp}
\end{equation} 
Here
\begin{equation}
\tau(\eta) \equiv \int_\eta^{\eta_0} \dot\tau(\eta') d\eta'
\end{equation}
is the optical depth between $\eta$ and the present.
The combination $\dot\tau e^{-\tau}$ is the visibility function
and expresses the probability that a photon last scattered between
$d\eta$ of $\eta$ and hence is sharply peaked at the last scattering
epoch.

Similarly, the polarization solutions follow from the radial 
decomposition of the 
\begin{equation}
-\sqrt{6} \dot\tau P^{(m)} \left[ {}_{2}^{\vphantom{2}} G_2^m {\bf M}_+ 
	      +{}_{-2}^{\vphantom{2}} G_2^m {\bf M}_- \right]	
\end{equation} 
source.  From Eqn.~(\ref{eqn:ebparity}), the solutions,
\begin{eqnarray}
{E^{(m)}_\ell(\eta_0,k) \over 2\ell+1} &=&  -\sqrt{6}
\int_0^{\eta_0} d\eta \, \dot\tau e^{-\tau} P^{(m)}_{\vphantom{\ell}}
(\eta)\epsilon_\ell^{(m)}(k(\eta_0-\eta)) \, ,\nonumber\\
{B^{(m)}_\ell(\eta_0,k) \over 2\ell+1} &=&  -\sqrt{6} 
\int_0^{\eta_0} d\eta \, \dot\tau e^{-\tau} P^{(m)}_{\vphantom{\ell}}
(\eta)\beta_\ell^{(m)}(k(\eta_0-\eta)) \, . 
\label{eqn:polsources}
\end{eqnarray}
immediately follow as well.

Thus the structures of $j_\ell^{(\ell'm)}$, $\epsilon_\ell^{(m)}$,
and $\beta_\ell^{(m)}$ shown in Figs. \ref{fig:radialtemp}
and \ref{fig:radialpol} directly reflects the angular power of
the sources $S_{\ell'}^{(m)}$ and $P^{(m)}$. 
There are several general results that can be read off the 
radial functions.  
Regardless of the source behavior in $k$, the 
$B$-parity polarization for scalars vanishes, dominates
by a factor of $6$ over the electric parity at $\ell \gg 2$
for the vectors, and is reduced by a factor of $8/13$
for the tensors at $\ell \gg 2$ [see Eqn.~(\ref{eqn:ebratio})].

Furthermore, 
the power spectra in $\ell$ can rise no faster than 
\begin{eqnarray}
\ell^2 C_\ell^{\Theta\Theta(m)}  \propto  \ell^{2+  2|m|}, \quad && \quad 
\ell^2 C_\ell^{EE(m)} \propto  \ell^{6 - 2|m|}, \nonumber \\
\ell^2 C_\ell^{BB(m)}  \propto  \ell^{6 - 2|m|}, \quad && \quad
\ell^2 C_\ell^{\Theta E(m)}  \propto \ell^4, 
\label{eqn:bleeding}
\end{eqnarray}
due to the aliasing of plane-wave power to $\ell \ll k(\eta_0-\eta)$
(see Eqn.~\ref{eqn:lowlaliasing})
which leads to interesting constraints on scalar temperature
fluctuations \cite{HuSug} and polarization 
fluctuations (see \S \ref{sec:seedcl}).  

Features in $k$-space
in the $\ell = |m|$ moment at {\it fixed} time 
 are increasingly well preserved in
$\ell$-space as $|m|$ increases,
but may be washed out if the source is not
well localized in time.  Only sources involving the
visibility function $\dot\tau e^{-\tau}$ are required to be
well localized at last scattering.  However even features
in such sources will be washed out if they occur
in the $\ell = |m|+1$ moment, such as the scalar dipole and the vector
quadrupole (see Fig.~\ref{fig:radialtemp}).
Similarly
features in the vector $E$ and tensor $B$ modes are washed out.

The geometric properties of the temperature-polarization cross 
power spectrum $C_\ell^{\Theta E}$ can also be read off the integral
solutions.  It is first instructive however to rewrite
the integral solutions as $(\ell \ge 2)$
\begin{eqnarray}
{\Theta_\ell^\se(\eta_0,k) \over 2\ell + 1}\, 
  &= &
\int_0^{\eta_0} d\eta \, e^{-\tau}
\left[(\dot \tau \Theta_0^\se +\dot \tau\Psi  + \dot\Psi - \dot\Phi )
	j_\ell^{(00)} + 
	\dot \tau v_B^\se j_\ell^{(10)}
        +\dot\tau P^\se  j_\ell^{(20)} \right], \nonumber\\
{\Theta_\ell^\ve(\eta_0,k) \over 2\ell+1} 
&=& \int_0^{\eta_0} d\eta
\, e^{-\tau} \left[ \dot\tau ( v_B^\ve - V) j^{(11)}_\ell 
+ (\dot\tau P^\ve + {1 \over \sqrt{3}} k V) j^{(21)}_\ell \right], \nonumber\\
{\Theta_\ell^\te(\eta_0,k) \over 2\ell+1} 
&=& \int_0^{\eta_0} d\eta
\, e^{-\tau} \left[ \dot\tau P^\te - \dot H \right] j_\ell^{(22)},
\label{eqn:explicitintegral}
\end{eqnarray}
where we have integrated the scalar and vector equations by 
parts noting that $d e^{-\tau} /d\eta = \dot\tau e^{-\tau}$.
Notice that $\Theta_0^\se+\Psi$ acts as an effective temperature
by accounting for the gravitational redshift from the potential
wells at last scattering.  
We shall see in \S\ref{sec:tight}
that $v_B^\ve \approx V$ at last scattering which 
suppresses the first term in the vector equation.   Moreover,
as discussed in \S \ref{sec:radial} and shown in 
Fig.~\ref{fig:radialcross}, the vector dipole terms ($j^{(11)}_\ell$)
do not correlate well with the polarization ($\epsilon^\ve_\ell$)
whereas the quadrupole terms ($j^{(21)}_\ell$) do.

The cross power spectrum contains two pieces: the relation between
the temperature and polarization sources $S_{\ell'}^{(m)}$ and
$P^{(m)}$ respectively  and the differences in
their projection as anisotropies on the sky.  The latter is 
independent of the model and provides interesting consequences
in conjunction with tight coupling and
causal constraints on the sources.  
In particular, the {\it sign} of the correlation is determined by
\cite{CrossNote}
\begin{eqnarray}
{\rm sgn}[C_\ell^{\Theta E(0)}] &=& 
-{\rm sgn}[P^{(0)}(\Theta_0^\se+\Psi)] \, ,\nonumber\\
{\rm sgn}[C_\ell^{\Theta E(1)}] &=& 
-{\rm sgn}[P^{(1)}(\sqrt{3}\dot\tau P^{(1)} + kV)]\, , \nonumber\\
{\rm sgn}[C_\ell^{\Theta E(2)}] &=& \hphantom{-}
{\rm sgn}[P^{(2)}(\dot\tau P^{(2)} - \dot H)] \, ,
\label{eqn:crosssign}
\end{eqnarray}
where the sources are evaluated at last scattering and we have
assumed that $|\Theta_0^\se + \Psi|  \gg |P^\se|$ 
as is the case for standard recombination (see \S 
\ref{sec:tight}).    The scalar Doppler effect couples
only weakly to the polarization due to differences
in the projection (see \S \ref{sec:radial}). 
The important aspect is that relative to the sources, the tensor
cross spectrum has an opposite sign due to the projection 
(see Fig.~\ref{fig:radialcross}).

These integral solutions are also useful in calculations.  
For example, they may be employed with approximate solutions to the sources
in the tight coupling regime to gain physical insight
on anisotropy formation (see \S \ref{sec:tight} and 
\cite{HuSug,ZalHar}). 
Seljak \& Zaldarriaga
\cite{SelZalInt} have obtained exact solutions through numerically
tracking the evolution of the source by solving
the truncated  Boltzmann hierarchy equations.   Our expression
agree with \cite{SelZal,KamKosSte,SelZalInt} where they overlap.

\section{Photon-Baryon Fluid} \label{sec:tight}

Before recombination, Thomson scattering between the photons and
electrons and coulomb interactions between the electrons and
baryons were sufficiently rapid that the photon-baryon system
behaves as a single tightly coupled fluid.  Formally, one
expands the evolution equations in powers of the
Thomson mean free path over the wavelength and horizon
scale.
Here we briefly review well known results for the scalars (see e.g. 
\cite{PeeYu,Kai}) to show how vector or vorticity perturbations
differ in their behavior (\S \ref{sec:tightfluid}).  
In particular, the lack of 
pressure support for the vorticity changes the relation between
the CMB and metric fluctuations. 
We then study the higher order effects of 
shear viscosity and polarization generation
from scalar, vector and tensor perturbations (\S \ref{sec:viscosity}).
We identify signatures in the temperature-polarization power spectra 
that can help separate the types of perturbations.  
Entropy generation and heat 
conduction only occur for the scalars (\S \ref{sec:entropy})
and leads to differences in the dissipation rate for
fluctuations (\S \ref{sec:dissipation}). 

\subsection{Compression and Vorticity} \label{sec:tightfluid}

For the $(m=0)$ scalars, the well-known result of expanding
the Boltzmann equations (\ref{eqn:boltz}) for $\ell=0,1$ 
and the baryon
Euler equation (\ref{eqn:sbaryon}) is 
\begin{eqnarray}
\dot \Theta_0^\se &=& 
	- {k \over 3} \Theta_1^\se - \dot\Phi  \, ,\nonumber\\
(m_{\rm eff} \Theta_1^\se)\dot{\vphantom{A}} &=& 
k(\Theta_0^\se + m_{\rm eff}\Psi)  \, ,
\label{eqn:tcconteul}
\end{eqnarray}
which represent the photon fluid continuity and Euler equations
and gives the baryon fluid quantities directly as
\begin{equation}
\dot \delta_B  =  {1 \over 3}\dot\Theta_0^\se   , \qquad
v_B^\se  =  \Theta_1^\se  ,
\end{equation}
to lowest order.  Here $m_{\rm eff}= 1+R$ where recall that $R$ is
the baryon-photon momentum density ratio.  
We have dropped the viscosity term
$\Theta_2^\se = {\cal O}(k/\dot\tau)\Theta_1^\se$ (see \S 
\ref{sec:viscosity}).  
The effect of the baryons is to introduce a Compton drag term that
slows the oscillation and
enhances infall into gravitational potential wells $\Psi$. That these 
equations describe forced acoustic oscillations in the fluid is
clear when we rewrite the equations as
\begin{equation}
(m_{\rm eff} \dot\Theta_0^\se)\dot{\vphantom{A}} + {k^2 \over 3}\Theta_0^\se
= -{k^2 \over 3} m_{\rm eff}\Psi - (m_{\rm eff}\dot\Phi)\dot{\vphantom{A}} \, ,
\label{eqn:oscillator}
\end{equation}
whose solution in the absence of metric fluctuations
is
\begin{eqnarray}
\Theta_0^\se &=& A m_{\rm eff}^{-1/4} \cos(ks + \phi)\, , \nonumber\\
\Theta_1^\se &=& \sqrt{3} A m_{\rm eff}^{-3/4}\sin(ks + \phi) \, ,
\label{eqn:oscillatorsoln}
\end{eqnarray}
where $s=\int c_{\gamma B} d\eta = \int (3m_{\rm eff})^{-1/2} d\eta$ 
is the sound horizon, $A$ is
a constant amplitude and $\phi$ is
a constant phase shift.
In the presence of potential perturbations,
the redshift a photon experiences
climbing out of a potential well makes the effective temperature 
$\Theta_0^\se + \Psi$ (see Eqn.~\ref{eqn:explicitintegral}), 
which satisfies
\begin{equation}
[m_{\rm eff} (\dot\Theta_0^\se +\dot\Psi)]\,\dot{\vphantom{A}} + 
{k^2 \over 3}(\Theta_0^\se +\Psi)
= -{k^2 \over 3} R\Psi + 
	[m_{\rm eff}(\dot\Psi - \dot\Phi)]\,\dot{\vphantom{A}} ,
\label{eqn:effectivetemp}
\end{equation}
and shows that the effective force on the oscillator is due
to baryon drag $R\Psi$ and differential gravitational redshifts
from the time dependence of the metric.  
As seen in Eqn.~(\ref{eqn:explicitintegral}) 
and (\ref{eqn:oscillatorsoln}),
the effective temperature
at last scattering forms the main contribution at last scattering
with the Doppler effect $v_B^\se = \Theta_1^\se$ 
playing a secondary role for $m_{\rm eff} > 1$.  Furthermore, 
because of the nature
of the monopole versus dipole projection, features in $\ell$
space are mainly created by the effective temperature
(see Fig.~\ref{fig:radialtemp} and \S \ref{sec:integral}).

If
$R \ll 1$, then one expects contributions of 
${\cal O}(\ddot \Psi - \ddot \Phi)/k^2$ to the oscillations
in $\Theta_0^\se +\Psi$ in addition to the initial fluctuations.
These acoustic contributions should be compared with the
${\cal O}(\Delta\Psi-\Delta\Phi)$ contributions from gravitational
redshifts in a time dependent metric after last scattering.
The stimulation of oscillations at $k\eta \gg 1$ thus either
requires large or rapidly varying metric fluctuations.  In 
the case of the former, acoustic oscillations would be small
compared to gravitational redshift contributions.

Vector perturbations on the other hand lack pressure support
and cannot generate acoustic
or compressional waves.  The
tight coupling expansion of the photon $(\ell=1,m=1)$ 
and baryon Euler
equations (\ref{eqn:boltz}) and (\ref{eqn:vbaryon}) 
leads to 
\begin{equation}
[ m_{\rm eff} (\Theta^\ve_1 - V) ]\dot{\hphantom{A}} = 0 \, ,
\label{eqn:vectortightcoupling}
\end{equation}
and $v^\ve_B = \Theta^\ve_1$.  Thus the vorticity in the 
photon baryon fluid is of equal amplitude to the vector 
metric perturbation.  In the absence of sources, it is
constant in a photon-dominated fluid and decays as
$a^{-1}$ with the expansion in a baryon-dominated fluid.
In the presence of sources, the solution
is
\begin{equation}
\Theta^{(1)}_1(\eta,k) = V(\eta,k) + {1 \over m_{\rm eff}}
        [\Theta^{(1)}(0,k)-V(0,k)] \, ,
\label{eqn:vectortcsoln}
\end{equation}
so that the photon dipole tracks the evolution of the metric
fluctuation. With $v_B^\ve=\Theta^\ve_1$ in 
Eqn.~(\ref{eqn:tempsources}), vorticity
leads to a Doppler effect in the CMB of magnitude
on order the vector metric fluctuation at last scattering $V$ in
contrast to scalar acoustic effects which depend on the time
rate of change of the metric.   

In turn the vector metric
depends on the vector anisotropic stress of the matter
as
\begin{equation}
V(\eta_*,k) = - 8\pi G a_*^{-2} \int_0^{\eta_*} 
	d\eta a^4 (p_f \pi_f^\ve + \pi_s^\ve)/k .
\end{equation}
In the absence of sources $V \propto a^{-2}$ and decays with
the expansion.  They are thus generally negligible if the universe
contains only the usual fluids. Only seeded models such as cosmological
defects may have their contributions to
the CMB anisotropy dominated by vector modes. 
However even though the vector to scalar {\it fluid}
contribution to the anisotropy for seeded models is of order
$k^2 V/(\ddot\Psi -\ddot\Phi)$ and may be large, 
the vector to scalar
gravitational redshift contributions, of order 
$V/(\Psi-\Phi)$ is not necessarily large.  Furthermore from
the integral solution for the vectors
Eqn.~(\ref{eqn:explicitintegral}) and the tight coupling
approximation Eqn.~(\ref{eqn:vectortcsoln}), the fluid effects
tend to cancel part of the gravitational effect.

\begin{figure}[t]
\begin{center}
\leavevmode
\epsfxsize=4in \epsfbox{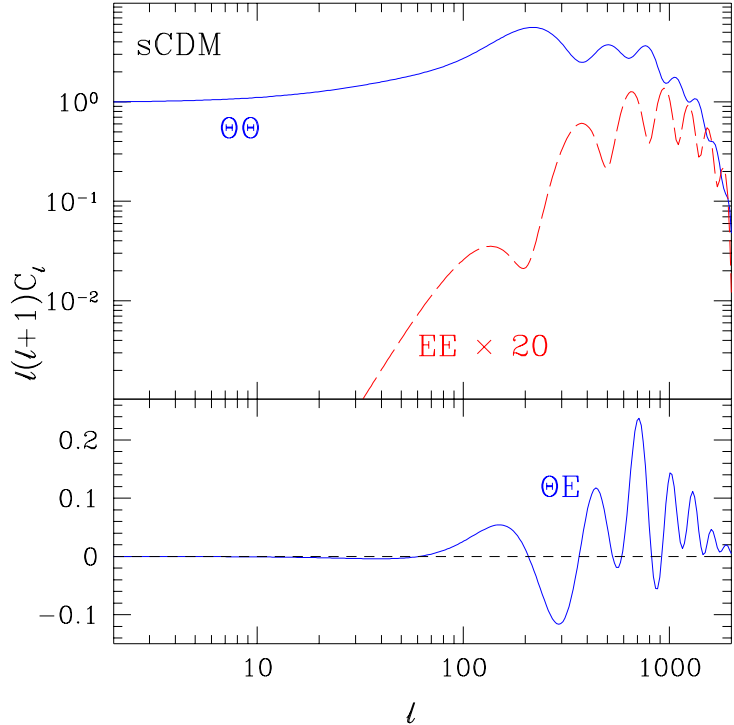} 
\end{center}
\caption{Power spectra for the standard cold dark matter
model (scale invariant scalar adiabatic initial conditions with 
$\Omega_0=1$, $h=0.5$ and $\Omega_B h^2=0.0125$).  Notice that
$B$-polarization is absent, $E$-polarization scales as $\ell^4$ 
at large angles and the cross correlation ($\Theta E$)
is negative at large
angles and reflects the acoustic oscillations at small angles.
In particular the phase of the $EE$ and $\Theta E$
acoustic peaks is set by the temperature oscillations  $\Theta\Theta$
(see \S \ref{sec:viscosity}).}
\label{fig:scdm}
\end{figure}

\subsection{Viscosity and Polarization} \label{sec:viscosity}

Anisotropic stress represents shear viscosity in the fluid
and is generated as tight coupling breaks down on small scales
where the photon diffusion length is comparable to the
wavelength.  
For the photons, anisotropic stress is related 
to the quadrupole moments of
the distribution $\Theta_2^{(m)}$ which is in turn coupled
to the E-parity polarization $E_2^{(m)}$.  The zeroth order
expansion of the polarization $(\ell=2)$ equations 
(Eqn.~\ref{eqn:epol})
gives
\begin{equation}
E_2^{(m)} = -{\sqrt{6}\over 4} \Theta_2^{(m)}, \qquad
B_2^{(m)} = 0
\label{eqn:Etightcoupling}
\end{equation}
or $P^{(m)} = {1 \over 4} \Theta_2^{(m)}$.
The quadrupole $(\ell=2)$ component of the temperature hierarchy
(Eqn.~\ref{eqn:boltz})
then becomes to lowest order in
$k/\dot\tau$
\begin{equation}
\Theta_2^{(m)} = {4 \over 9}\sqrt{4-m^2} {k \over \dot\tau} 
	\Theta_1^{(m)}, \qquad
P^{(m)} = {1 \over 9}\sqrt{4-m^2} {k \over \dot\tau} 
\Theta_1^{(m)}\, ,
\label{eqn:quadrupoletightcoupling}
\end{equation}
for scalars and vectors.   In the tight coupling limit, the
scalar and vector sources of polarization traces 
the structure of the photon-baryon fluid velocity.
For the tensors,
\begin{equation}
\Theta_2^{(2)} = -{4 \over 3}{ \dot H \over \dot \tau}\, ,\qquad
P^{(2)} = -{1 \over 3}{ \dot H \over \dot \tau} \, .
\label{eqn:tpoltc}
\end{equation}
Combining 
Eqns.~(\ref{eqn:Etightcoupling}) and 
(\ref{eqn:quadrupoletightcoupling}), 
we see that polarization fluctuations
are generally suppressed with respect to metric or
temperature fluctuations. 
They are proportional to the quadrupole moments in the temperature
which are suppressed by scattering.  Only as the optical depth
decreases can polarization be generated by scattering.  Yet then
the fraction of photons affected also decreases.  In the standard
cold dark matter model, the polarization is less than $5\%$ of
the temperature anisotropy at its peak (see Fig.~\ref{fig:scdm}).  

These  scaling relations 
between the metric and anisotropic 
scattering sources of the temperature and polarization are important
for understanding the large angle behavior of the polarization and
temperature polarization cross spectrum.  Here last scattering
is effectively instantaneous compared with the scale of the 
perturbation and the tight coupling remains a good approximation
through last scattering.

For the scalars, the Euler equation (\ref{eqn:tcconteul})
may be used to express the scalar velocity and hence the
polarization in terms of
the effective temperature,
\begin{equation}
\Theta_1^\se = m_{\rm eff}^{-1} \int k(\Theta_0^\se + m_{\rm eff}\Psi)
d\eta \, .
\end{equation}
Since $m_{\rm eff} \sim 1$, 
$\Theta_1^\se$ has the same sign as $\Theta_0^\se + \Psi$
before $\Theta_0^\se + \Psi$ itself can change
signs,
assuming reasonable initial conditions. 
It then follows that $P^{(0)}$ is also of the same sign
and is of order
\begin{equation}
P^{(0)} \sim (k\eta){k \over \dot\tau} [\Theta_0^\se + \Psi]\, ,
\label{eqn:spoltc} 
\end{equation}
which is strongly suppressed for $k\eta \ll 1$.  
The definite sign leads to a definite prediction
for the sign of the temperature polarization cross correlation
on large angles.   

For the vectors
\begin{equation}
P^{(1)} = {\sqrt{3} \over 9} {k \over \dot\tau} V \, ,
\label{eqn:vpoltc}
\end{equation}
and is both suppressed and has a definite sign 
in relation to the metric fluctuation.   The tensor relation
to the metric is given in Eqn.~(\ref{eqn:tpoltc}). 
In fact, in all three cases the dominant
source of temperature perturbations has the {\it same} sign as
the anisotropic scattering source $P^{(m)}$. From
Eqn.~(\ref{eqn:crosssign}), differences in the
sense of the 
cross correlation between temperature and polarization thus
arise only due to geometric reasons in the projection 
of the sources
(see~Fig.~\ref{fig:radialcross}).  On angles larger than the 
horizon at last scattering, the scalar and vector $C_\ell^{\Theta E}$
is negative whereas the tensor cross power is positive 
\cite{CriCouTur,CrossNote}.  

On smaller scales, the scalar polarization follows the velocity
in the tight coupling regime.  It is instructive to recall the
solutions for the acoustic oscillations from 
Eqn.~(\ref{eqn:oscillatorsoln}).  The velocity oscillates $\pi/2$
out of phase with the temperature and hence the $E$-polarization
acoustic peaks will be out of phase with the temperature
peaks (see Fig.~\ref{fig:scdm}).  The cross correlation oscillates
as $\cos(ks +\phi)\sin(ks +\phi)$ and hence has twice the 
frequency. Thus between peaks of the polarization and temperature
power spectra (which represents {\it both} peaks and troughs of
the temperature amplitude) the cross correlation 
peaks.  The structure of the cross correlation
can be used to measure the acoustic phase $\phi$ ($\phi \approx 0$
for adiabatic models) and how
it changes with scale just as the temperature but like the $EE$
power spectrum \cite{HuSpeWhi} has the added 
benefit of probing slightly larger scales than
the first temperature peak.  
This property can help distinguish adiabatic
and isocurvature models due to causal constraints on 
the generation of acoustic waves at the horizon at last scattering
\cite{SpeZal}.

Finally, polarization also increases the viscosity of the fluid
by a factor of $6/5$, which has significant effects for the 
temperature.
Even though the viscous imperfections of the fluid
are small in the tight coupling region they can lead to significant
dissipation of the fluctuations over time 
(see \S \ref{sec:dissipation}).

\subsection{Entropy and Heat Conduction} \label{sec:entropy}

Differences in the bulk velocities of the photons and baryons
$\Theta_1^{(m)} - v_B^{(m)}$ also represent imperfections in the
fluid that lead to entropy generation and heat conduction.
The baryon Euler equations (\ref{eqn:sbaryon}) and (\ref{eqn:vbaryon})
give
\begin{eqnarray}
\Theta_1^\se - v_B^\se  & = & {R \over \dot\tau}\left[
	\dot v_B^\se + {\dot a \over a} v_B^\se - k\Psi \right]  \,,
	\nonumber\\
\Theta_1^\ve - v_B^\ve  & = & {R \over \dot\tau}\left[
	\dot v_B^\ve - \dot V + {\dot a \over a} (v_B^\ve - V)
	\right] \, ,
\label{eqn:slippage}
\end{eqnarray}
which may be iterated to the desired order in $1/\dot\tau$.
For scalar fluctuations, this slippage 
leads to the generation of non-adiabatic
pressure or entropy fluctuations
\begin{eqnarray}
\Gamma_{\gamma B} &\equiv & (\delta p_{\gamma B} 
	- c_{\gamma B}^2 \delta \rho_{\gamma B})/p_{\gamma B} \nonumber\\
 & = & - {4 \over 3}{R \over 1+R} \int (\Theta_1^\se - v_B^\se) 
	k\, d\eta \, ,
\end{eqnarray}
as the local number density of baryons to photons changes.  
Equivalently, this can be viewed as heat conduction in the fluid.
For vorticity fluctuations, these processes do not occur since there
are no density, pressure, or temperature differentials in the fluid. 


\subsection{Dissipation} \label{sec:dissipation}

The generation of viscosity and heat conduction in the fluid
dissipates fluctuations through the Euler equations
with (\ref{eqn:quadrupoletightcoupling}) 
and (\ref{eqn:slippage}),  
\begin{eqnarray}
(1+R)\dot\Theta_1^\se &=& k \left[ \Theta_0^\se  + (1+R) \Psi 
		\right] + {k \over \dot \tau} R^2 \dot \Theta_0^\se 
		- {16 \over 45} {k^2 \over \dot \tau}\Theta_1^\se 
		\, , \nonumber\\
(1+R)\dot\Theta_1^\ve &=& (1+R) \dot V - {k \over \dot \tau} 
	{4 \over 15}k\Theta_1^\ve \, , 
\label{eqn:tighteuler}
\end{eqnarray}
where we have dropped the $\dot a/a$ factors under the assumption
that the expansion can be neglected
during the dissipation period.  We have also employed 
Eqn.~(\ref{eqn:vectortightcoupling}) to 
eliminate higher order terms in the
vector equation.  With the continuity equation for the
scalars $\dot \Theta_0^{(0)} = -k \Theta_1^{(0)}/3 - \dot\Phi$
(see Eqn.~\ref{eqn:boltz} $\ell=0$, $m=0$ ), we obtain
\begin{eqnarray}
\ddot \Theta_0^\se + {1 \over 3} {k^2 \over \dot\tau}
        \left[ {R^2 \over (1+R)^2} + {16 \over 15}{1 \over 1+R}
        \right] \dot \Theta_0^\se + {k^2 \over 3(1+R)}\Theta_0^\se = 
	-{k^3 \over 3}\Psi - \ddot \Phi \, ,
\label{eqn:dampedoscillator}
\end{eqnarray}
which is a damped forced oscillator equation.

An interesting case to consider is the behavior in the absence
of metric fluctuations $\Psi$, $\Phi$, and $V$.  The result,
immediately apparent from Eqn.~(\ref{eqn:tighteuler}) and
(\ref{eqn:dampedoscillator}), is that 
the acoustic amplitude $(m=0)$ and vorticity $(m=1)$ damp as 
$\exp[-(k/k_D^{(m)})^2]$ where
\begin{eqnarray}
\left[{1 \over k_D^\se } \right]^2
	&=& {1 \over 6} \int d\eta \, { 1 \over \dot \tau}
        {{R^2 + 16(1+R)/15}\over {(1+R)^2}} \, ,\nonumber\\
\left[{1 \over k_D^\ve } \right]^2
        &=&  {4 \over 15} \int d\eta{ 1\over \dot \tau}
{1 \over {1 + R} } \, .
\label{eqn:diffusionlength}
\end{eqnarray}
Notice that dissipation is less rapid for the vectors
compared with the scalars once
the fluid becomes baryon dominated $R\gg 1$ because of the
absence of heat conduction damping.  In principle, this allows
vectors to contribute more CMB anisotropies at small scales
due to fluid contributions.  In practice, the dissipative cut off
scales are not very far apart since $R \simlt 1$ at recombination.

Vectors may also dominate if there is a continual metric
source.  There is a competition between the metric source
and dissipational sinks in Eqns.~(\ref{eqn:tighteuler}) and
(\ref{eqn:dampedoscillator}).  Scalars retain contributions
to $\Theta^\se_0 + \Psi$ of 
${\cal O}(R\Psi,(\ddot\Psi-\ddot\Phi)/k^2)$ 
(see Eqn.~(\ref{eqn:effectivetemp}) and \cite{HuWhiDamp}).  
The vector solution becomes 
\begin{equation}
\Theta_1^\ve(\eta) = e^{-[k/k_D^\ve(\eta)]^2}
	\int_0^\eta d\eta' \dot V
		e^{[k/{k_D^\ve}(\eta')]^2}  ,
\end{equation}
which says that if variations in the metric are rapid compared with
the damping then $\Theta_1^\ve = V$ and damping does not occur.  

\section{Scaling Stress Seeds} \label{sec:seed}

Stress seeds provide an interesting example of
the processes considered above by which scalar, vector, and
tensor metric perturbations are generated and affect
the temperature and polarization of the CMB.   They are also the
means by which cosmological defect models form structure
in the universe.  As  part of the class of isocurvature models,
all metric
fluctuations, including the (scalar) curvature perturbation
are absent in the initial conditions.
To explore the basic properties
of these processes, we employ simple examples
of stress seeds under the restrictions they are causal and scale
with the horizon length. Realistic
defect models may be constructed by superimposing such
simple sources in principle.

We begin by discussing the form of the stress seeds themselves
(\S \ref{sec:seedsource})
and then trace the processes by which they form metric perturbations
(\S \ref{sec:seedmetric})
and hence CMB anisotropies (\S \ref{sec:seedcl}).

\subsection{Causal Anisotropic Stress} \label{sec:seedsource}

Stress perturbations are fundamental to seeded models
of structure formation because causality combined
with energy-momentum conservation forbids 
perturbations in the energy or momentum density until matter
has had the opportunity to move around inside the horizon (see
e.g.~\cite{VeeSte}).
Isotropic stress, or pressure, only arises for scalar perturbations
and have been considered in detail by \cite{HuSpeWhi}.  Anisotropic
stress perturbations can also come in vector and tensor types
and it is their effect that we wish to study here.  Combined
they cover the full range of possibilities available to 
causally seeded models such as defects.

\begin{figure}[t]
\begin{center}
\leavevmode
\epsfxsize=3.5in \epsfbox{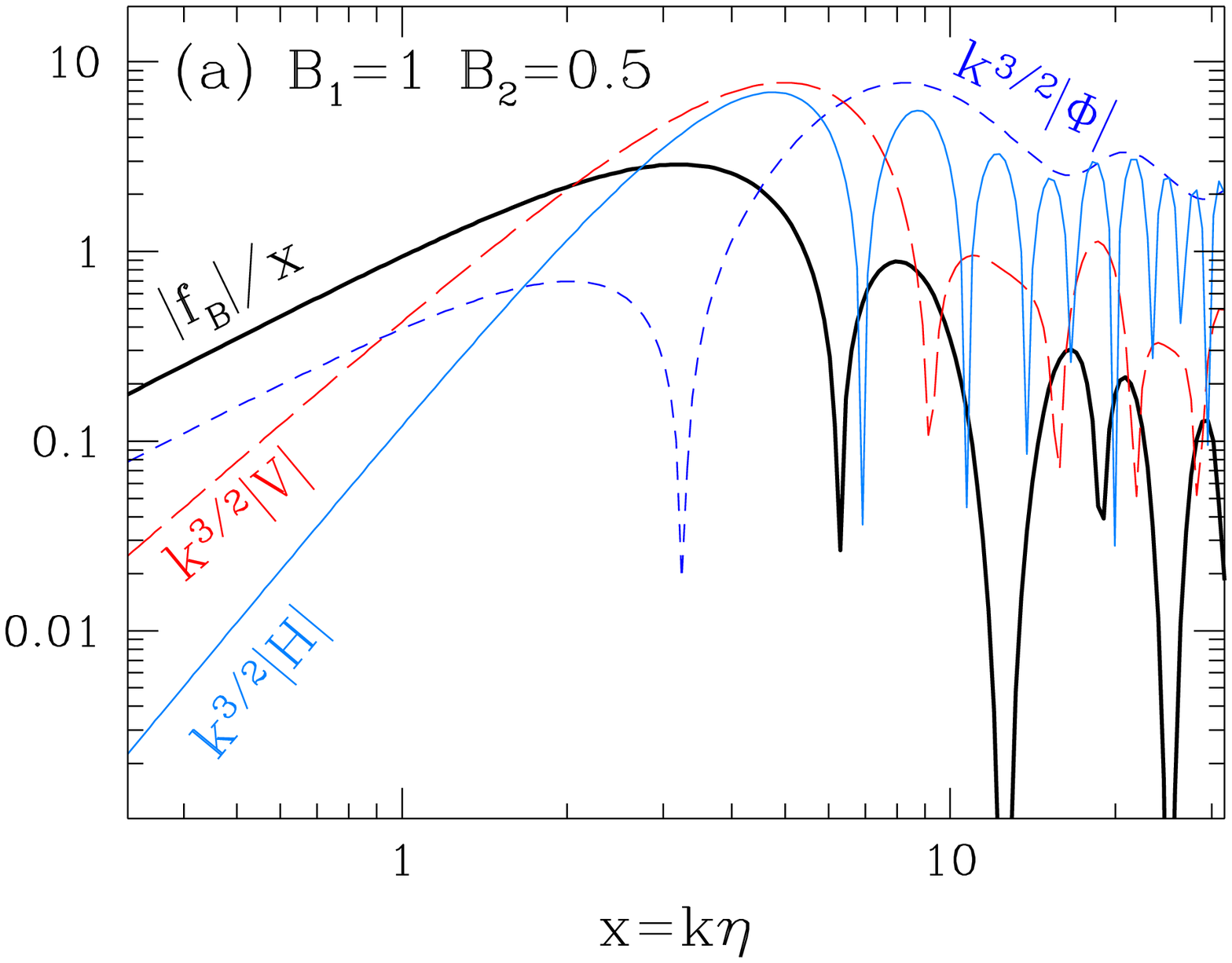} 
\epsfxsize=3.5in \epsfbox{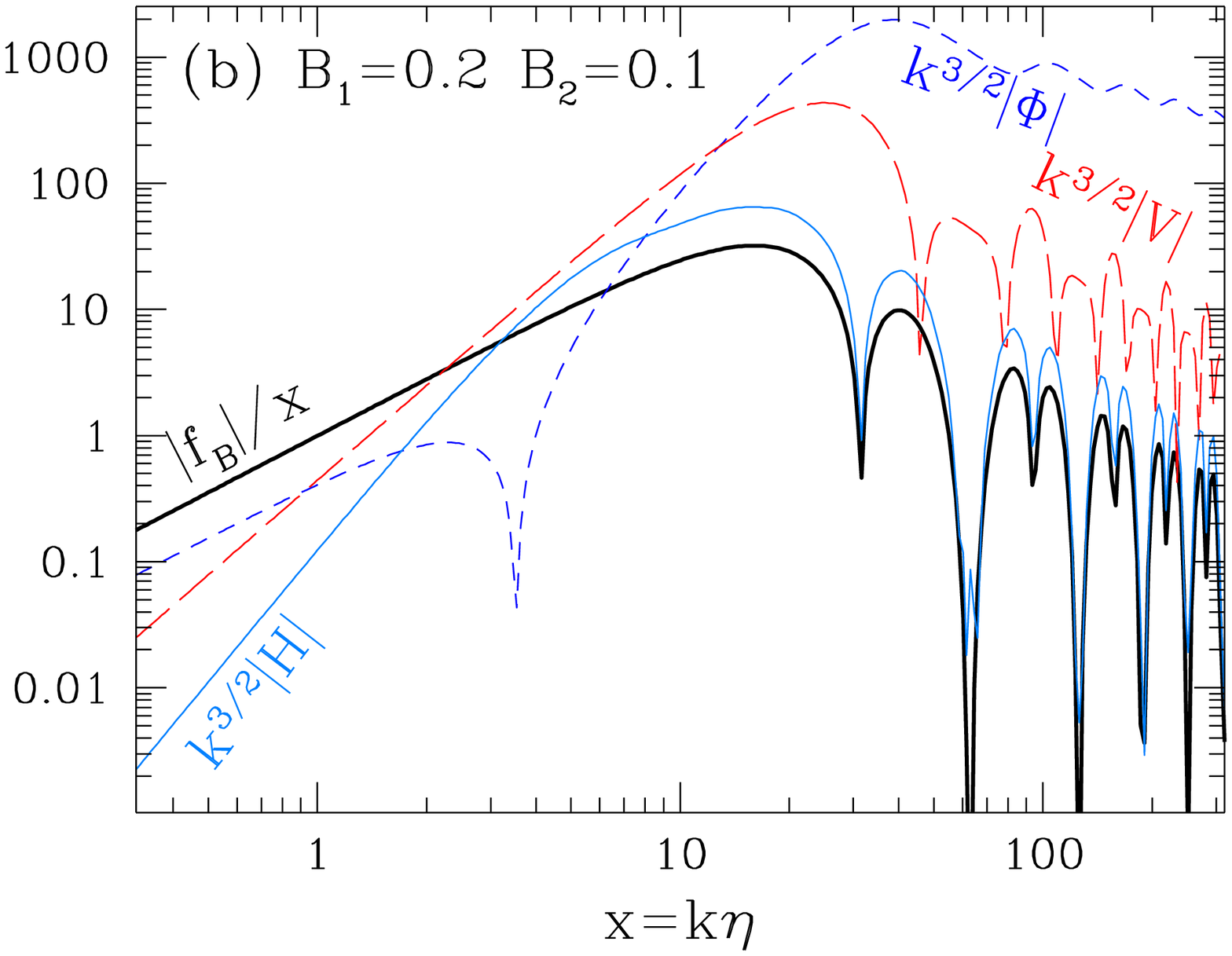} 
\end{center}
\caption{Metric fluctuations from scaling anisotropic stress seeds
sources.  The same anisotropic stress seed (bold solid lines
$a^2 \pi_s \propto f_B/x$) produce
qualitatively different scalar (short-dashed), vector (long-dashed), 
and tensor (solid) metric perturbations.  As discussed in the
text the behavior scales with the characteristic time of
the source $x_c \propto B_1^{-1}$.  The left panel (a) shows a source
which begins to decay as soon as causally permitted ($B_1=1$)
and the right panel (b) the effect of delaying the decay 
($B_1=0.2$).  We have displayed the results here for a 
photon-dominated universe for simplicity.}
\label{fig:pi}
\end{figure}

We impose two constraints on the anisotropic stress seeds: causality
and scaling.  Causality implies that correlations in the stresses
must vanish outside the horizon.  
Anisotropic stresses represent spatial derivatives of
the momentum density
and hence vanish as $k^2$ for $k\eta \ll 1$.
Scaling requires that the
fundamental scale is set by the current horizon so that evolutionary
effects are a function of $x=k\eta$.   A convenient form
that satisfies these criteria is \cite{HuSpeWhi,turoknew}
\begin{equation}
4\pi G a^2 \pi_s^{(m)} = A^{(m)}\eta^{-1/2} f_B(x) \, ,
\end{equation}
with
\begin{equation}
f_B^{}(x) = {6 \over B_2^{2} - B_1^{2}}
              \left[ {\sin(B_1 x) \over (B_1 x)}
              -{\sin(B_2 x) \over (B_2 x)} \right] \, ,
\label{eqn:basispi}
\end{equation}
with $0 < (B_1,B_2) < 1$.  We caution the reader that though
convenient and complete, this choice of basis is not 
optimal for representing the currently popular set of 
defect models.
It suffices for our purposes here since we only wish to 
illustrate general properties of the anisotropy formation process.

Assuming $B_1 > B_2$, $B_1$ controls the characteristic
time after horizon crossing that the stresses 
are generated, i.e. the peak in $f_B$
scales as $k\eta_c \equiv x_c \propto B_1^{-1}$ 
(see Fig.~\ref{fig:pi}).  
$B_2$ controls the rate of decline
of the source at late times.
In the general case, the seed may be a sum of different pairs 
of $(B_1,B_2)$ which may also differ between scalar, vector, 
and tensor components.  

\begin{figure}[t]
\begin{center}
\leavevmode
\epsfxsize=4in \epsfbox{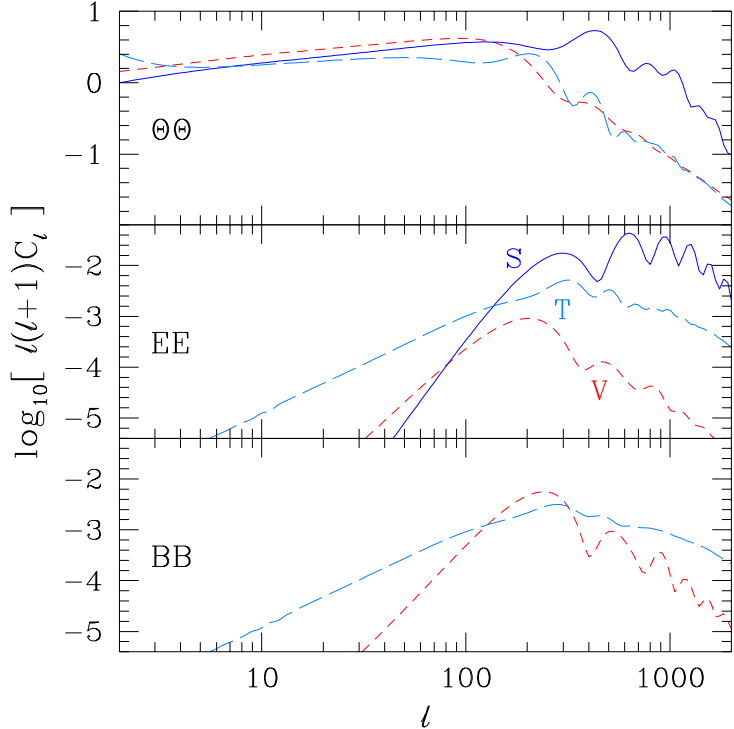} 
\end{center}
\caption{Temperature and polarization power spectra for
a scaling anisotropic stress seeds with the minimal characteristic
time $B_1=1$ for scalars (S, solid), vectors (V, short-dashed), and
tensors (T, long-dashed).  
Scalar temperature fluctuations at
intermediate scales are dominated by acoustic contributions which
then damp at small scales.  $B$-parity polarization 
contributions are absent for the scalars, larger by an
order of magnitude than $E$-parity contributions for the vectors
and similar to but smaller than the $E$-parity for the tensors.  
Features in the vector and tensor spectra are artifacts of
our choice of source and are unlikely to be present in 
a realistic model.  The background cosmology is set to $\Omega_0=1$,
$h=0.5$, $\Omega_B h^2 = 0.0125$.
}
\label{fig:b1}
\end{figure}

\subsection{Metric Fluctuations} \label{sec:seedmetric}

Let us consider how the anisotropic stress seed sources generate
scalar, vector and tensor metric fluctuations.   The form
of Eqn.~(\ref{eqn:basispi}) implies that the metric perturbations
also scale so that $k^3| h |^2 = f(x=k\eta)$ where $f$ may be
different functions for $h = (\Psi,\Phi,V,H)$.
Thus scaling in the defect field also implies scaling for
the metric evolution and consequently the purely
{\it gravitational} effects
in the CMB as we shall see in the next section.  Scattering
introduces another fundamental scale, the horizon at last
scattering $\eta_*$, which we shall see  breaks the scaling in
the CMB.

It is interesting to consider differences in the evolutions
for the same anisotropic stress seed, $A^{(m)}=1$ with 
$B_1$ and $B_2$ set equal for the scalars, vectors and tensors.  
The basic tendencies
can be seen by considering the  behavior at early times 
$x \simlt  x_c$.  If $x \ll 1$ as well,
then the contributions
to the metric fluctuations scale as
\begin{eqnarray}
k^{3/2}\Phi/f_B  =  {\cal O}(x^{-1}) \, , \qquad & & \qquad
k^{3/2}\Psi/f_B  =  {\cal O}(x^{-1}) \, , \nonumber\\
k^{3/2} V/f_B    =  {\cal O}(x^{0})  \, , \qquad & & \qquad
k^{3/2} H/f_B    =  {\cal O}(x^{1}) \, ,
\end{eqnarray}
where $f_B = x^2$ for $x \ll 1$. Note that the sources
of the scalar fluctuations in this limit are the anisotropic
stress and momentum density rather than energy density
(see Eqn.~\ref{eqn:Poisson}).   
This behavior is displayed
in Fig.~\ref{fig:pi}(a).  For the scalar and tensor evolution,
the horizon scale enters in separately from the characteristic
time $x_c$.  For the scalars, the stresses move matter around 
and generate density fluctuations as $\rho_s \sim x^2 \pi_s$.
The result is that the evolution of $\Psi$ and $\Phi$ steepens by
$x^2$ between $1 \simlt x \simlt x_c$. 
For the tensors, the 
equation of motion takes the form of a damped driven oscillator
and whose amplitude follows the source.  Thus the tensor
scaling becomes shallower in this regime.  For $x \simgt x_c$
both the source and the metric fluctuations decay.  Thus
the {\it maximum} metric fluctuation scales as
\begin{eqnarray}
k^{3/2}\Phi/f_B(x_c)  =  {\cal O}(x_c^{1})\, , \qquad && \qquad
k^{3/2}\Psi/f_B(x_c)  =  {\cal O}(x_c^{1})\, ,  \nonumber\\
k^{3/2} V/f_B(x_c)    =  {\cal O}(x_c^{0} )\, , \qquad && \qquad
k^{3/2} H/f_B(x_c)    =  {\cal O}(x_c^{-1} ). 
\label{eqn:maxmetric}
\end{eqnarray}
For a late characteristic time $x_c > 1$, fluctuations in the
scalars are larger than vectors or tensors for the same
source (see Fig.~\ref{fig:pi}b).  The ratio of acoustic to 
gravitational
redshift contributions from the scalars scale
as $x_c^{-2}$ by virtue of pressure support in 
Eqn.~(\ref{eqn:effectivetemp})
and thus acoustic oscillations 
become subdominant as $B_1$ decreases.

\subsection{CMB Anisotropies} \label{sec:seedcl}

Anisotropy and structure formation in causally seeded, or in
fact any isocurvature model, proceeds by a qualitatively different 
route than the conventional
adiabatic inflationary picture.  As we have seen, fluctuations
in the metric are only generated inside the horizon rather than
at the initial conditions (see \S \ref{sec:seedmetric}).  
Since CMB anisotropies probe scales
outside the horizon at last scattering, one would hope that
this striking difference can be seen in the CMB.  Unfortunately,
gravitational redshifts between last scattering and today 
masks the signature in the temperature anisotropy.
The scaling
ansatz for the sources described in \S \ref{sec:seedsource} 
in fact leads to near scale invariance
in the large angle temperature because fluctuations are
stimulated in the same way for each $k$-mode as it crosses the horizon
between last scattering and the present.  While these 
models generically leave a different signature in modes
which cross the horizon before last scattering \cite{HuSug,HuSpeWhi}, 
models which mimic adiabatic inflationary predictions can
be constructed \cite{turoknew}.  

\begin{figure}[t]
\begin{center}
\leavevmode
\epsfxsize=4in \epsfbox{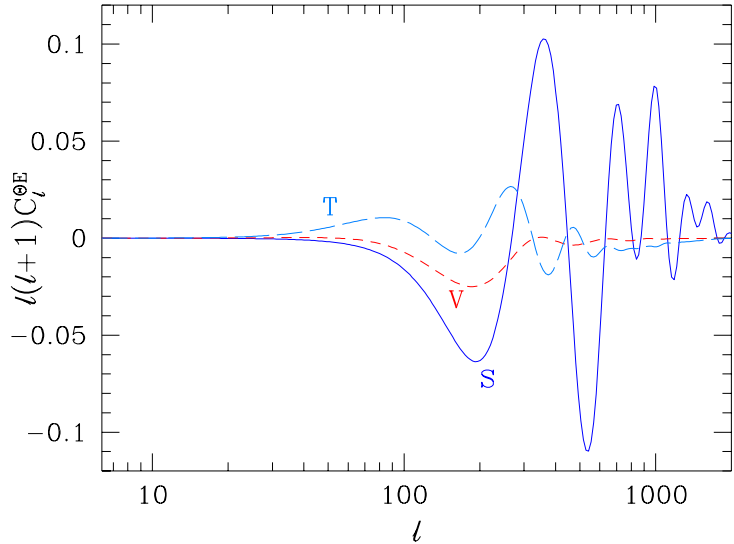} 
\end{center}
\caption{Temperature-polarization cross power spectrum for the model
of Fig.~\ref{fig:b1}. 
Independent of the nature
of the sources, the cross power at angles larger than that subtended
by the horizon at last scattering  
is negative for the scalars
and vectors and positive for the tensors.  The more complex structure
for the scalars at small angular  scales reflects the correlation
between the acoustic effective temperature  and velocity at
last scattering.
} 
\label{fig:te}
\end{figure}

Polarization provides a more direct
test in that it can only be generated
through scattering.  The large angle polarization reflects fluctuations
near the horizon at last scattering and so may provide a direct
window on such causal, non-inflationary models of structure
formation.  One must be careful however to separate scalar,
vector and tensor modes whose different large angle behaviors
may obscure the issue. 
Let us now illustrate these considerations with the specific examples 
introduced in the
last section.

The metric fluctuations produced by the seed sources 
generate CMB anisotropies through the Boltzmann
equation
(\ref{eqn:boltz}).
We display an example with $B_1=1$ and
$B_2=0.5$ in Fig.~\ref{fig:b1}.  Notice that 
scaling in the sources does indeed lead to near scale invariance
in the large angle temperature but not the large angle polarization.
The small rise toward the quadrupole for the tensor temperature
is due to
the contribution of long-wavelength gravity waves that are 
currently being generated and depends on how rapidly they
are generated after horizon crossing.
Inside the horizon at last scattering (here $\ell \simgt 200$),
scalar fluctuations generate acoustic waves as discussed in
\S \ref{sec:tight} which dominate for small characteristic
times $x_c \approx 1$.  
On the other hand, these contributions are strongly
damped below the thickness of the last scattering surface
by dissipational processes.  Note that features in the vector
and tensor spectrum shown here are artifacts of our choice of
source function.  In a realistic model, the superposition of
many sources of this type will wash out such features.  The
general tendencies however do not depend on the detailed form
of the source.  Note that vector and tensor contributions 
damp more slowly and hence may contribute significantly
to the small-angle temperature anisotropy. 

\begin{figure}[t]
\begin{center}
\leavevmode
\epsfxsize=4in \epsfbox{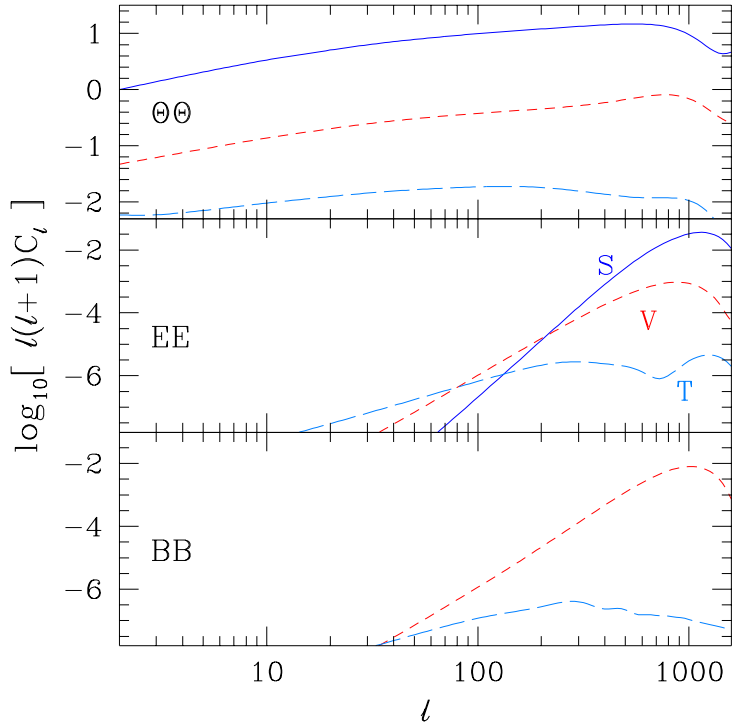} 
\end{center}
\caption{Same as Fig.~\ref{fig:b1} except with a larger
characteristic time $B_1 =0.2$, $B_2=0.1$.  Scalar gravitational
redshift effects now dominate over scalar acoustic as well
as vector and tensor contributions for the same stress source
due the process by which stress perturbations generate
metric fluctuations (see Fig.~\ref{fig:pi}).}  
\label{fig:b02}
\end{figure}

Polarization can only be generated by scattering of a quadrupole
temperature anisotropy. 
For seeded models, scales outside the horizon
at last scattering  $k\eta_* \simlt 1$ 
have not formed significant metric fluctuations
(see e.g. Fig.~\ref{fig:pi}).  Hence quadrupole fluctuations,
generated from the metric fluctuations through 
Eqns.~(\ref{eqn:tpoltc}),
(\ref{eqn:spoltc}), and (\ref{eqn:vpoltc}), are also suppressed.
The power in $k$ of the polarization thus drops sharply
below
$k\eta_* = 1$.  This drop of course corresponds to a lack of
large angle power in the polarization.  However its form at 
low $\ell$ depends on geometric aspects of the projection from
$k$ to $\ell$.   In these models, the {\it large} 
angle polarization is dominated
by projection aliasing of power from {\it small} scales $k\eta_*
\simgt 1$.  The asymptotic expressions of Eqn.~(\ref{eqn:bleeding})
thus determine the large angle behavior of the polarization:
$\ell^2 C_\ell \propto \ell^6$ for scalars $(EE)$, 
$\ell^4$ for vectors
and $\ell^2$ for tensors ($EE$ and $BB$); the cross spectrum 
$(\Theta E)$ goes as $\ell^4$ for each contribution.
For comparison, the scale invariant
adiabatic
inflationary prediction has scalar
polarization ($EE$) dropping off as $\ell^4$ and cross
spectrum ($\Theta E$) as $\ell^2$ 
from Eqn.~(\ref{eqn:spoltc}) because of the constant
potential above the horizon (see Fig.~\ref{fig:scdm}).
Seeded models thus predict a more rapid reduction in the 
{\it scalar} polarization
for the same background cosmology.

Polarization can also help separate the three types of fluctuations.
In accord with the general prediction (see \S \ref{sec:integral}),
scalars produce no $B$-parity polarization, whereas 
vector $B$-parity dominates $E$-parity polarization 
by a factor of $6$ and tensor $B$-parity is suppressed
by a factor of $8/13$ (see Eqn.~(\ref{eqn:ebratio})). 
Differences also arise in the temperature-polarization
cross power spectra $C_\ell^{\Theta E}$ shown in Fig.~\ref{fig:te}.  
Independent of the nature of the source, above the angle the
horizon subtends at last scattering, scalar and vector temperature
perturbations from the last scattering surface \cite{CrossNote}
are anticorrelated with polarization, whereas they are correlated for
tensor perturbations (see \S \ref{sec:viscosity}
and \cite{CriCouTur}).  Inside
the horizon, the scalar polarization follows the scalar
velocity which is $\pi/2$ out of phase with the effective
temperature (see Eqn.~\ref{eqn:quadrupoletightcoupling}).
In the adiabatic model, scalar cross correlation reverses 
signs before the first acoustic peak, as compression overcomes
the gravitational redshift of the Sachs-Wolfe effect, unlike the 
isocurvature models (see Fig.~\ref{fig:scdm} and \cite{SpeZal}). 
The sign test to
distinguish scalars from tensors must thus be performed on
scales larger than twice the first peak.  Conversely, to use
the cross correlation to distinguish adiabatic from isocurvature
fluctuations, the scalar and tensor contributions must be
separated.

How do these results change with the model for
the seeds?
As we increase the characteristic time $x_c$ by decreasing 
$B_1=0.2$, the main effect
comes from differences in the generation of metric fluctuations
discussed in \S \ref{sec:seedmetric}.  For the same amplitude
anisotropic stress, scalars contributions dominate the 
vector and tensor contributions by factors of $x_c = B_1^{-1}$
(see Eqn.~\ref{eqn:maxmetric}).  Note however that the scalar
contributions come from the gravitational redshifts between last
scattering and today rather than the acoustic oscillations 
(see \S \ref{sec:tightfluid}) and hence produce no strong features.
Because of the late generation of metric fluctuations in these
models, the peak in the polarization spectra is also shifted
with $x_c$.  Note however that the qualitative behavior
of the polarization described above remains the same.

Although these examples do not exhaust the full range of possibilities
for scaling seeded models, the general behavior is representative.
Equal amplitude anisotropic stress sources tend to produce 
similar large angle temperature anisotropies if the source is
active as soon as causally allowed $x_c \approx 1$.  Large
angle scalar polarization is reduced as compared with
adiabatic inflationary models because of causal constraints 
on their formation. 
This behavior is not as marked in vectors and tensors 
due to the projection geometry
but the relative amplitudes of the $E$-parity and $B$-parity 
polarization as well as the $\Theta E$ cross correlation
can be used to separate them independently 
of assumptions for the seed sources.  Of course in practice
these tests at large angles will be difficult to apply due
to the smallness of the expected signal.  

Reionization increases
the large angle polarization signal because the quadrupole anisotropies
that generate it can be much larger \cite{Rei}.  This  occurs since
decoupling occurs gradually and the scattering is no longer
rapid enough to suppress anisotropies.  The prospects for separating
the scalars, vectors and tensors based on polarization consequently
also improve \cite{ZalSpeSel}.

For angles smaller than
that subtended by the horizon at last scattering, the relative
contributions of these effects depends on a competition 
between scalar gravitational
and acoustic effects and the differences in the generation 
and damping behavior of the scalar, vector and tensor perturbations.

\section{Discussion}

We have provided a new technique for the study of temperature 
and polarization anisotropy formation in the CMB which
introduces a simple and systematic 
representation for their angular distributions.
The main virtue of this approach is that the
gravitational and scattering sources are directly related to
observable properties in the CMB.   One can then explore
properties that are {\it independent} of the source, which
tell us the broad framework, e.g.~the classical cosmological
parameters and the nature of fluctuations in the early universe,
and identify properties that are {\it dependent} on the source,
which can help pin down the model for structure formation.   
An example of the former is the fact that scalar fluctuations
generate no magnetic parity polarization \cite{SelZal,KamKosSte},
vectors generate mainly magnetic parity polarization, 
and tensors generate comparable but somewhat smaller magnetic parity
polarization.   Large angle polarization of the three components
are also constrained by model-independent geometric arguments in
its slope and its correlation with the temperature anisotropy.
If the scalar contributions can be isolated from
the vectors, tensors and other foreground sources of polarization
from these and other means,
these constraints translate into a robust distinction 
between isocurvature and adiabatic models for structure formation.

In our representation, the temperature and polarization distributions 
are projections on the sky of four simple sources, 
the metric fluctuation (via the gravitational redshift), 
the intrinsic temperature at last scattering,
the baryon velocity at last scattering (via the Doppler effect)
and the temperature and 
polarization quadrupoles at last scattering (via the angular 
dependence of Compton scattering).  
As such, it better reveals the power of
the CMB to probe the nature of these sources and extract information
on the process of structure formation in the universe.  As an 
example, we have explored how general properties of scaling stress
seeds found in cosmological defect models manifest themselves in 
the temperature and polarization power spectra.  The framework we 
have provided here should be useful for determining the robust 
signatures of specific models for structure formation as well as the 
reconstruction of the true model for structure formation 
from the data as it becomes available.

\vskip 0.5truecm

\noindent{\it Acknowledgements:}  We would like to thank Uros
Seljak and Matias Zaldarriaga for many useful discussions. W.H.
was supported by the W.M. Keck Foundation.
\clearpage

\def\vertspm{$\vertsp$}
\clearpage

TAB-2. Commonly used symbols.  $m=0,\pm 1,\pm 2$ for the scalars, vectors, and
tensors. For the fluid variables $f \rightarrow \gamma$ for the photons,
$f \rightarrow B$ for the baryons and $\gamma B$ for the photon-baryon fluid.
$X = \Theta$, $E$, $B$ for the temperature-polarization power spectra. 
\begin{center}
\vskip 0.5truecm
\begin{tabular}{| c | l | c | }
\hline
\, Symbol \quad  & 
\quad\qquad\qquad Definition  & \, Eqn. \quad \\ \hline
$\Psi,\Phi$   & \quad Scalar metric & (\ref{eqn:smetric})
        \vertspm\ \\ \hline
$\Theta_\ell^{(m)}$ & \quad $\Delta T/T$ moments & (\ref{eqn:decomposition})
        \vertspm\ \\ \hline
$\alpha,\beta,\gamma$ &\quad Euler angles & (\ref{eqn:composition}) 
        \vertspm\ \\ \hline
$\beta^{(m)}_\ell,\epsilon^{(m)}_\ell$ 
	&\quad Radial $B,E$ function & (\ref{eqn:radialpol})
        \vertspm\ \\ \hline
$\delta_f $ &\quad Fluid density perturbation & (\ref{eqn:sstress})
        \vertspm\ \\ \hline
$\eta$ &\quad Conformal time & (\ref{eqn:genmetric})
        \vertspm\ \\ \hline
$\kap{s}{\ell}{m}$ &\quad Clebsch-Gordan coefficient & 
	(\ref{eqn:kapfactor})
	\vertspm\ \\ \hline
$\rho_f, \rho_s$ &\quad Fluid, seed density & (\ref{eqn:sstress}) 
        \vertspm\ \\ \hline
$\pi_f^{(m)},\pi_f^{(m)}$ &\quad Fluid, seed anisotropic stress & (\ref{eqn:sstress})
        \vertspm\ \\ \hline
$\theta,\phi$ &\quad Spherical coordinates in $\hat{k}$ frame\qquad
        & (\ref{eqn:temperaturebasis})
        \vertspm\ \\ \hline
$\tau$ &\quad Thomson optical depth & (\ref{eqn:collisionimplicit})
        \vertspm\ \\ \hline
$B_\ell^{(m)}$ &\quad $B$-pol. moments & (\ref{eqn:decomposition})
        \vertspm\ \\ \hline
$\vec{C} $ &\quad Collision term & (\ref{eqn:fullcollision})
        \vertspm\ \\ \hline
${C_\ell^{X\widetilde X(m)}} $ &\quad $X\widetilde X$ power spectrum from $m$ & (\ref{eqn:cl})
        \vertspm\ \\ \hline
$E_\ell^{(m)}$ &\quad $E$-pol. moments & (\ref{eqn:decomposition})
        \vertspm\ \\ \hline
$\vec{G} $ &\quad Gravitational redshift term & (\ref{eqn:gravred})
        \vertspm\ \\ \hline
$G_\ell^m$ &\quad Temperature basis & (\ref{eqn:temperaturebasis})
        \vertspm\ \\ \hline
${}_{\pm 2}^{\vphantom{2}} G_\ell^m$ &\quad Polarization basis & (\ref{eqn:polarizationbasis})
        \vertspm\ \\ \hline
$H$   &\quad Tensor metric & (\ref{eqn:tmetric})
        \vertspm\ \\ \hline
\end{tabular}
\qquad
\begin{tabular}{| c | l | c | }
\hline
\, Symbol \quad  & 
\quad\qquad\qquad Definition  & \, Eqn. \quad \\ \hline
${\bf M}_\pm $ &\quad (Pauli) matrix basis & (\ref{eqn:matrixbasis}) 
        \vertspm\ \\ \hline
$P^{(m)}$ &\quad Anisotropic scattering source & (\ref{eqn:polsource})
        \vertspm\ \\ \hline
$Q^{(0)}$ &\quad Scalar basis & (\ref{eqn:scalareigen})
        \vertspm\ \\ \hline
$Q^{(1)}_i$ &\quad Vector basis & (\ref{eqn:vectoreigen})
        \vertspm\ \\ \hline
$Q^{(2)}_{ij}$ &\quad Tensor basis & (\ref{eqn:tensoreigen})
        \vertspm\ \\ \hline
$R$   &\quad $B/\gamma$ momentum density & (\ref{eqn:Rdef})
        \vertspm\ \\ \hline
$S_\ell^{(m)}$ & \quad Temperature source  & (\ref{eqn:tempsources})
        \vertspm\ \\ \hline
$V$   &\quad Vector metric & (\ref{eqn:vmetric})
        \vertspm\ \\ \hline
\quad $\Spy{s}{\ell}{m}$ \quad  &\quad Spin-$s$ harmonics & \quad (\ref{eqn:spinbasis})\quad
        \vertspm\ \\ \hline
$j_{\ell}^{(\ell'm)}$ &\quad Radial temp function \quad & (\ref{eqn:jdef})
        \vertspm\ \\ \hline
$\vec{k}$ &\quad Wavenumber & (\ref{eqn:temperaturebasis})
        \vertspm\ \\ \hline
$k_D^{(m)}$ &\quad Damping wavenumber & (\ref{eqn:diffusionlength})
        \vertspm\ \\ \hline
$\ell$ &\quad Multipole & (\ref{eqn:spinbasis}) 
        \vertspm\ \\ \hline
$m_{\rm eff}$ &\quad Effective mass $1+R$ & (\ref{eqn:tcconteul})
        \vertspm\ \\ \hline
$\hat{n}$ &\quad Propagation direction & (\ref{eqn:planewave})
        \vertspm\ \\ \hline
$p_f, p_s$ &\quad Fluid, seed pressure & (\ref{eqn:sstress}) 
        \vertspm\ \\ \hline
$v_f^{(m)}$ &\quad Fluid velocity  & (\ref{eqn:sstress})
        \vertspm\ \\ \hline
$v_s^{(m)}$ &\quad Seed momentum density \quad& (\ref{eqn:sstress})
        \vertspm\ \\ \hline
$w_f$ &\quad $p_f/\rho_f$ & (\ref{eqn:sfluideqn})
        \vertspm\ \\ \hline
\end{tabular}
\end{center}


\vskip 1truein

{Published in: {\it Physical Review D} {\bf 56}, 596 (1997)
\vskip 0.5truein
{\tt whu@ias.edu}

{\tt http://www.sns.ias.edu/$\sim$whu}

{\tt http://www.sns.ias.edu/$\sim$whu/polar/polar.html}
and astro-ph/9706147 for a descriptive accounts of this work.


\begin{thebibliography}{99}
\bibitem{CrossNote} There can be a small contribution from unequal
time correlations at the very lowest $\ell$ if the main signal
from the last scattering surface is falling sufficiently rapidly.
\end{thebibliography}
\end{document}